\documentclass[useAMS,usenatbib]{mn2e}
\usepackage[pdf]{pstricks}
\usepackage{amsmath}
\usepackage[]{txfonts}
\def\simless{\mathbin{\lower 3pt\hbox
{$\rlap{\raise 5pt\hbox{$\char'074$}}\mathchar"7218$}}}   
\def\simmore{\mathbin{\lower 3pt\hbox
{$\rlap{\raise 5pt\hbox{$\char'076$}}\mathchar"7218$}}}   
\newcommand{\bh}{pe}
\newcommand{\eqb}{\begin{eqnarray}}
\newcommand{\eqe}{\end{eqnarray}}
\newcommand{\mel}{m_{\rm e}}
\newcommand{\mpr}{m_{\rm p}}
\newcommand{\Bcr}{B_{\rm cr}}
\newcommand{\gp}{\gamma_{\rm p}}
\newcommand{\gpmx}{\gamma_{\rm p, \max}}
\newcommand{\lpinj}{\ell_{\rm p}^{\rm inj}}
\newcommand{\leinj}{\ell_{\rm e}^{\rm inj}}
\newcommand{\af}{\alpha_{\rm f}}
\newcommand{\pg}{ p \pi}

\newcommand{\esco}{ \epsilon'_{\rm s}}
\newcommand{\es}{\epsilon_{\rm s}}
\newcommand{\emin}{\epsilon_{\min}}
\newcommand{\emax}{\epsilon_{\max}}
\newcommand{\xs}{x_{\rm s}}
\newcommand{\xth}{x_{\rm th}}
\newcommand{\rb}{r_{\rm b}}
\newcommand{\fpg}{f_{\rm p\pi}}
\newcommand{\fbh}{f_{\bh}}
\newcommand{\tpg}{t_{\rm p \pi}}
\newcommand{\tbh}{t_{\bh}}
\newcommand{\tcr}{t_{\rm cr}}
\newcommand{\sth}{\sigma_{\rm T}}
\newcommand{\tgg}{\tau_{\gamma \gamma}}
\usepackage{graphicx}
\usepackage{epsfig} 
\usepackage{epstopdf}

\title[Bethe-Heitler emission in BL~Lacs]
{Bethe-Heitler emission in BL Lacs: filling the gap between X-rays
and $\gamma$-rays}

\author[M. Petropoulou \& A. Mastichiadis]
{M. Petropoulou$^{1}$\thanks{E-mail: mpetropo@purdue.edu} \thanks{Einstein Postdoctoral Fellow} \& A. Mastichiadis$^{2}$\thanks{E-mail: amastich@phys.uoa.gr}\\
$^{1}$Department of Physics and Astronomy, Purdue University, 525 Northwestern
Avenue, West Lafayette, IN 47907, USA\\
$^{2}$Department of Physics, University of Athens, Panepistimiopolis, GR 15783 Zografos, Greece}
\begin{document}
\date{Received / Accepted}
\pagerange{\pageref{firstpage}--\pageref{lastpage}} \pubyear{2014}

\maketitle

\label{firstpage}

\begin{abstract}
We present the spectral signatures of the Bethe-Heitler pair production ($\bh$) process
on the spectral energy distribution (SED) of blazars, in scenarios
where the hard $\gamma$-ray emission is of photohadronic origin.
If relativistic protons interact with the synchrotron blazar
photons producing $\gamma$ rays through photopion processes, we show
that, besides the $\sim 2-20$~PeV neutrino emission,
the typical blazar SED should have an emission feature due to the synchrotron emission of $\bh$ secondaries
that bridges the gap betweeen the low-and high-energy humps of the SED, namely in the energy range $40$~keV -- $40$~MeV.
We first present analytical expressions for the photopion and $\bh$
loss rates in terms of observable quantities of blazar emission. 
For the  $\bh$ loss rate in particular,
we derive a new approximate analytical expression for the case of
a power-law photon distribution, which has an excellent accuracy with the 
 numerically calculated exact one, 
especially at energies above the threshold for pair production. 
We show that for typical
blazar parameters, the photopair synchrotron emission
emerges in the hard X-ray/soft $\gamma$-ray energy range 
with a characteristic spectral shape and non negligible flux, 
which may be even comparable to the  hard
$\gamma$-ray flux  produced through photopion
processes. We argue that the expected ``$\bh$ bumps'' 
are a natural consequence of 
leptohadronic models,  and as such, they may indicate that
blazars with a three-hump SED are possible
emitters of high-energy neutrinos.
\end{abstract} 
  

\section{Introduction}
Blazars are a subclass of active galactic nuclei with a non-thermal continuum 
emission spanning many orders of magnitude in energy, i.e. 
from radio frequencies
up to high-energy $\gamma$-rays.
Their spectral energy distribution (SED) has a double humped appearance (e.g. \citealt{ulrichetal97, fossatietal98}) with
a broad low-energy component extending from radio up to UV, or in some extreme cases to $\gtrsim 1$~keV X-rays \citep{costamante01},
and the high-energy one covering the X-ray and $\gamma$-ray energy regime with a peak energy around
$0.1$ TeV, but this is not always clear (see e.g. \citealt{abdo11} for Mrk~421).
Although they exhibit variability in almost all
frequencies (e.g. \citealt{raiteri12}), the rapid high-energy variability is one of their striking features. 
Variability timescales may range between $\sim $ day and several hours \citep{kataoka01, sobolewska14}, while in some extreme cases they may 
reach down to a few minutes in GeV and TeV energies \citep{aharonian07, aleksic11, forschini13}. Their variable emission
when combined with the large inferred isotropic luminosities
provides strong evidence that blazar emission originates 
in relativistic jets that are closely aligned with our line of sight.

It is generally accepted that the low-energy component is the result of synchrotron radiation 
from relativistic electrons in the jet, yet the origin of the high-energy component remains an open issue.
Theoretical models for blazar emission are divided into leptonic and
leptohadronic, according to the type of particles responsible
for the high-energy emission. It is noteworthy that both have been successfully applied to blazars
(for a review, see \citealt{boettcher10}).
 In  pure leptonic scenarios, the high-energy component 
is the result of  inverse Compton scattering of electrons
in a photon field. As seed photons can serve the synchrotron photons produced by the same electron population
(SSC models: \citealt{maraschietal92, bloommarscher96, mastkirk97,konopelko03}) or/and photons  
from an external region (EC models),  such as the accretion disk 
\citep{dermeretal92,dermerschlickeiser93} or the broad line
region (BLR) \citep{sikoraetal94, ghisellinimadau96, boettcherdermer98}. EC models in particular,
are more relevant for flat spectrum radio quasars (FSRQs) and/or low-peaked blazars (LBLs)\footnote{
BL~Lac objects can be further divided in three categories according
to the peak frequency $\nu_{\rm s}$ of the low-energy component: low-frequency peaked (LBLs) for
 $\nu_{\rm s}\lesssim 10^{14}$~Hz, intermediate-frequency peaked (IBLs) for $ 10^{14}$~Hz$<\nu_{\rm s} \lesssim 10^{15}$~Hz,
 and high-frequency peaked (HBLs) if $\nu_{\rm s} > 10^{15}$~Hz.}; for modelling of specific sources belonging to both classes, see \cite{boettcherreimer13}.

In principle, both protons and electrons can be accelerated to relativistic energies (\citealt{biermannstrittmatter87, sironi13, globus14} and
references, therein). This argument consists the basis of leptohadronic models, where
the low-energy component of the SED is still explained by electron synchrotron radiation,
but the high-energy emission is now explained in terms of relativistic proton interactions in the jet.
Models that invoked interactions of relativistic protons with ambient matter (gas)
through $pp$ collisions (e.g. \citealt{stecker91, beall99, schuster02}) required high densities of the gas to explain the observed luminosities.
Attention was then drawn to proton interactions with low energy photons (photohadronic interactions), 
since their density in astrophysical environments usually exceeds that of the gas. In these models, 
target photons may be provided either externally from the jet (same as in EC models), i.e.
from the accretion disk \citep{bednarekprotheroe99} and the BLR \citep{atoyandermer01}, or they can 
be internally produced by the co-accelerated electrons. The similarity to the SSC models
is again obvious. The high-energy component can be 
the result of (i) the emission from an electromagnetic (EM) cascade 
initiated by the absorption of very high-energy (VHE) 
$\gamma$ rays produced through the $\pg$ process \citep{mannheim91, mannheimbiermann92, mannheim93}; (ii) 
synchrotron radiation of secondary pairs produced by the decays of charged pions \citep{petromast12, mastetal13};
(iii) neutral pion decay \citep{sahu13,cao14}; or proton synchrotron radiation,
for high enough magnetic fields \citep{aharonian00,  mueckeprotheroe01,
mueckeetal03,petropoulou14}.

Photohadronic interactions are comprised of two processes of astrophysical interest:
\begin{itemize}
 \item Bethe-Heitler  pair production ($\bh$)
 \eqb
 p + \gamma \rightarrow e^{+} + e^{-}.
 \eqe
 \item Photopion production  ($\pg$)
 \eqb
 p + \gamma  & \rightarrow & \pi^0 + p \\
 \pi^0  & \rightarrow & \gamma+\gamma \nonumber
 \eqe
 or
 \eqb
 p + \gamma & \rightarrow & \pi^{\pm} + n\left(\Delta^{++}\right), \\ 
 \pi^{\pm} & \rightarrow & \mu^{\pm} + \nu_{\mu}(\overline{\nu}_{\mu}), \nonumber \\
 \mu^{\pm} & \rightarrow & e^{\pm}+ \overline{\nu}_{\mu}(\nu_{\mu}) + \nu_{\rm e}(\overline{\nu}_{\rm e}) \nonumber .
 \eqe
\end{itemize} 
 Bethe-Heitler pair production is an often overlooked process since
it is not related with neutrino and neutron production, which is of particular interest
to high-energy astrophysics 
and a natural outcome of $\pg$ interactions \citep{sikoraetal87, kirkmast89, 
begelmanrudak90, giovanonikazanas90, waxmanbahcall97, atoyandermer01, atoyandermer03}.
Moreover, it is considered to be a subdominant proton cooling process, at least  
for protons that satisfy the threshold condition for $\pg$ production
 (e.g. \citealt{sikoraetal87}) and thus, is often neglected 
 in models of blazar emission (e.g. \citealt{cerruti11, boettcherreimer13, weidingerspanier13}). 
However, as we shall show in the present study, $\bh$ secondaries
are injected at different energies than photopion ones. Therefore,
even in cases where $\bh$ is subdominant, it can still leave a radiative
signature on the blazar spectrum.
An illustration of the $\bh$ contribution to
the injection of secondaries can be 
found in Fig.~8 of \cite{dmpr12}--henceforth DMPR12, and \cite{petropoulou14a}.

The aim of the present work is to study in more detail
the contribution of pairs injected by the $\bh$  process
to the SED of blazars. We focus on BL~Lac objects, a subclass of blazars named after the prototype object BL~Lacartae \citep{schmitt68} because
of two reasons: the majority of BL~Lac objects is detected in high-energy $\gamma$ rays ($\gtrsim 100$~GeV)
and is characterized by an extreme weakness of emission lines in the optical spectra, which 
suggests that any external radiation fields play a subdominant role in the formation of their spectra.
BL~Lacs, as less ``contaminated'' sources than FSRQs,  consistute
a more suitable class of objects for studying the emission signatures of photohadronic interactions.
The role of the  $\bh$  process in the emission of FSRQs will be the subject of a future work.

The theoretical framework that we adopt is described as follows: 
the low-energy emission is explained by synchrotron radiation of
relativistic electrons, whereas the observed high-energy (GeV-TeV) emission is
the result of synchrotron radiation from pairs produced by charged pion
decays. Pions in their turn, are the by-product of $\pg$ interactions
of co-accelerated protons with the internally produced synchrotron photons. We 
restrict our analysis to cases where
the high-energy spectra are not (severely) modified by EM cascades, which allows us to 
identify the observed $\gamma$-ray emission as the emission from the $\pg$ component.
Thus, in this framework, the $\gamma$-ray  blazar emission  is directly associated with 
neutrino emission at energies $\sim 2-20$~PeV, which may be of particular interest in the light of the recent
detection of astrophysical high-energy neutrinos \citep{icecube13, aartsen14}.

Our work is structured as follows. We begin in \S\ref{model}
with a description of our model. In \S\ref{analytical} we derive
analytical expressions for the typical energies
of synchrotron photons emitted by secondary pairs produced through the
$\bh$  and $\pg$ processes, and compare the respective proton
cooling rates for typical blazar parameters.
In \S\ref{numerical} we back up our analytical predictions with 
numerical examples. We discuss our results in \S\ref{discussion}
and conclude in \S\ref{summary} with a summary.

\section{The model}
\label{model}
We assume that the region responsible for the 
blazar emission can be described as a spherical blob of size $\rb$ 
that contains a tangled magnetic field of strength $B$, moving with a Doppler factor $\delta = \Gamma^{-1} \left(1-\beta \cos\theta \right)^{-1}$, where
$\Gamma$ is the bulk Lorentz factor and $\theta$ is the angle between the line of sight and the jet axis. We also assume
that both protons and primary electrons are injected uniformely with a constant rate into
the emission region, after having been accelerated 
to relativistic energies; their distribution is described by a power-law
with index $s_i$ and high-energy cutoff $\gamma_{\rm i, \max}$, where the subscript $i$
is used to discriminate between protons ($p$) and electrons ($e$).
Electrons lose energy through the synchrotron and inverse Compton processes, while
synchrotron radiation and photohadronic interactions count to the
main energy loss processes for relativistic protons. In the present context, we assume
that the target photons for the photohadronic interactions, which include both 
 $\bh$ 
and photopion production processes, are internally (or locally)
produced, i.e. they are the result of  
primary electron (and proton) synchrotron radiation.
Photohadronic interactions eventually
lead to {an increase of the lepton number density} in the emission region, since
both  $\bh$  and $\pg$ processes result in the injection of secondary pairs.
Thus, synchrotron and inverse Compton radiation of secondary pairs is an inevitable
outcome of photohadronic interactions, which, depending on the parameters,
may be imprinted on the blazar SED. 

\begin{figure}
 \centering
 \includegraphics[width=0.4\textwidth]{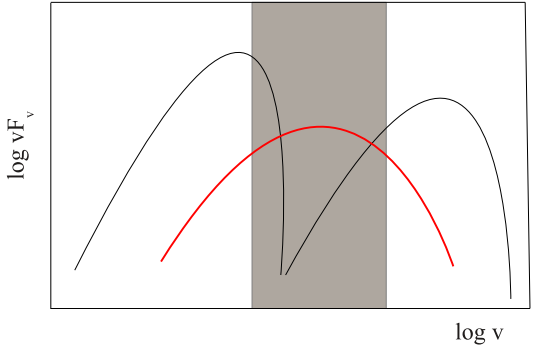}
 \caption{Sketch of a typical blazar SED in $\nu F_{\nu}$ units with the low and high energy humps
 shown with black lines. Synchrotron emission from Bethe-Heitler pairs
 is expected to appear as a third bump (red line) in the hard X-ray/soft $\gamma$-ray regime (grey colored zone).
 }
 \label{fig0}
\end{figure}

In the present work we investigate 
 a typical leptohadronic scenario that relates the
observed blazar $\gamma$-ray emission with a high-energy neutrino signal. 
In this context, the synchrotron emission from primary electrons and from the $\pg$ process contributes
   to the low and the high parts of the spectrum respectively. On the
   other hand, the  $\bh$  process typically injects pairs with lower energy than the
   $\pg$ secondaries, and we thus expect their synchrotron radiation to emerge
   between the two spectral humps. This is exemplified in 
   Fig.~\ref{fig0} where a fiducial blazar SED is shown schematically.
Besides the low-and high-energy humps (black lines), the contribution 
of the $\bh$ pairs to the SED is shown as a third component (red line) that 
emerges between the two (grey colored region).
We argue that the  $\bh$  emission that bridges the two humps of the SED
is a robust prediction of this model. In the following sections 
we investigate this argument in detail
using both analytical and numerical means.

\section{Analytical estimates}
\label{analytical}
When one fixes the primary
   electron and proton population so they can explain the characteristic
   double humps observed in blazars, then the $\bh$  component is determined
   automatically without the use of any additional parameter.
We present analytical expressions
for the typical synchrotron photon energies emitted by
secondary  $\bh$  and $\pg$ pairs, which we express in terms
of observable quantities of blazar emission.
We continue with the calculation of the $\bh$ and photopion energy loss rates
for monoenergetic and power-law target photon fields, and compare them
for parameter values typical for blazar emission.
We note that the analysis that follows is valid as long as 
the photon spectra are not modified by electromagnetic cascades
initiated by internal photon-photon absorption (see e.g.~\citealt{mannheim91}).
\subsection{Characteristic energies}
\label{A1}
We  estimate the characteristic energies of synchrotron photons emitted by secondary electrons, which
are the products of $\bh$ pair production and charged pion decay, and express them 
using observable quantities, such as the peak frequencies of the low-and high-energy
humps of the blazar SED. We list below the basic relations we used and note that 
primed variables denote quantities measured in the comoving frame of the region, whereas
unprimed variables are used for quantities measured in the 
observer's frame:
\begin{enumerate}
 \item 
 The low-energy bump of the SED is explained in terms
 of primary electron synchrotron radiation.  Its peak frequency
 $\nu_{\rm s}$ is then written as
 $\nu_{\rm s}= \delta (1+z)^{-1} h^{-1} (B/\Bcr) \mel c^2 \gamma_e^2$, 
 where 
  $z$ is the redshift of the source, $\Bcr=4.4\times 10^{13}$~G and $\gamma_e$
is the Lorentz factor of primary injected electrons. Using $10^{16}$~Hz as a typical value for the peak frequency we derive
the first relation
\eqb
\delta B \gamma_e^{2} = 3.5 \times 10^9 \nu_{\rm s, 16}.
\label{eq0}
\eqe
\item 
The proton threshold energy for $\pg$ interactions with 
the synchrotron photons of energy $\esco = h \nu_{\rm s} (1+z)/\delta$ is
given by 
\eqb
\gamma_{\rm p, \pg}^{\rm (th)} \simeq \frac{\xth}{\xs},
\label{eq1}
\eqe
where  $\xth= m_{\pi}/\mel$  for single pion production,
$m_{\pi} \simeq 145$~MeV/$c^2$ and $\xs = \esco / \mel c^2$ or 
\eqb
\xs = 8 \times 10^{-5} (1+z) \nu_{\rm s, 16} \delta^{-1},
\label{eq2}
\eqe
in terms of the obsverved peak frequency.
Inserting the above expression into eq.~(\ref{eq1}) we find
\eqb
\gamma_{\rm p, \pg}^{\rm (th)} \simeq  3.5 \times 10^6 (1+z)^{-1} \delta \nu_{\rm s, 16}^{-1}.
\label{eq3}
\eqe
\item 
The proton threshold energy for  $\bh$ pair production on synchrotron photons of energy $\esco$ is given by
\eqb
\gamma_{\rm p,   \bh }^{\rm (th)} \simeq 1.2 \times 10^4 (1+z)^{-1} \delta \nu_{\rm s, 16}^{-1},
\label{eq4}
\eqe
which is lower than the respective one for pion production by a factor of $\mel / m_{\pi}$.
\item 
We assume that the secondary electrons produced through $\pg$ interactions
from parent protons having energy $\gamma_{\rm p, \pg}^{\rm (th)}$
emit synchrotron radiation at high-energy $\gamma$ rays, 
e.g. $\nu_{\gamma} \gtrsim 10^{25}$~Hz. 
The secondary electrons are produced roughly with a Lorentz factor
$\gamma_{\rm e}^{\pg} \simeq \kappa_{\rm p} \gamma_{\rm p, \pg}^{\rm (th)} \mpr /4 \mel$, where $\kappa_{\rm p} \simeq 0.2$
is the mean inelasticity of $\pg$ interactions. Using eq.~(\ref{eq3}) we find the constraint
\eqb
\delta^3 B \simeq 30 (1+z)^3 \nu_{\rm s, 16}^2 \nu_{\gamma, 25}.
\label{eq5}
\eqe
\end{enumerate}
Finally, the typical  Lorentz factor of primary electrons is derived by combining eqs.~(\ref{eq0}) and (\ref{eq5})
\eqb
\gamma_{\rm e} \simeq 10^3 \delta (1+z)^{-1} \left(\nu_{\rm s, 16} \nu_{\gamma, 25} \right)^{-1/2}
\label{eq6}.
\eqe
Before continuing to the calculation of the proton energy loss rates due to photohadronic processes, 
it is useful to have an estimate of the energy range where the synchrotron emission from
$\bh$ pairs emerges. 

For proton-photon collisions taking place close to the threshold, 
the maximum Lorentz factor of the produced pairs
is $\gamma_{\rm e}^{\rm   \bh } \approx \gamma_{\rm p}$ \citep{mastkirk95,kelneraharonian08}. Using the expressions (\ref{eq4}) and (\ref{eq5})
we find 
\eqb
\nu_{\rm s}^{\rm  \bh } \simeq 5.2 \times 10^{16} \nu_{\gamma, 25} \ {\rm Hz},
\label{eq7}
\eqe
which falls in the UV/soft X-ray energy band. 
However,  $\bh$  emission at these energies should not have any observable effect on the SED, since
the injection rate of   $\bh$  pairs close to the threshold is small. This is a direct outcome of the
fact that the product of 
the  $\bh$ cross section and proton inelasticity has its maximum at $\gamma_{\rm p} x \approx 16$, where $x$ is the energy
of an arbitrary photon in $\mel c^2$ units (see e.g. Fig.~1 in \citealt{mastetal05}).
For a power-law proton distribution, which is the case under consideration here, protons with 
energies above the photopair threshold, i.e. $\gamma_{\rm p} \gtrsim \gamma_{\rm p,  \bh }^{\rm (th)}$,
will also interact with  photons of energy $\esco$. In this case, however, the interactions occur away from the threshold.
For $\bar{\epsilon} = 2 \gamma_{\rm p} \esco \gg \mel c^2$, where $\bar{\epsilon}$ is the energy of the synchrotron photons as measured
in the proton's rest frame, the pairs acquire a higher maximum Lorentz factor given 
by $\gamma_{\rm e}^{\rm  \bh } \approx 4 \gamma_{\rm p}^2 \xs \gtrsim 
4 \left(\gamma_{\rm p,  \bh }^{\rm (th)}\right)^2 \xs$ \citep{kelneraharonian08}.
Using eqs.~(\ref{eq2}), (\ref{eq4}) and (\ref{eq5}) we find that 
\eqb
\nu_{\rm s}^{\rm  \bh } \gtrsim 3.3 \times 10^{18}  \nu_{\gamma, 25} \ {\rm Hz},
\label{eq8}
\eqe
which corresponds to the hard X-ray/soft $\gamma$-ray regime, where
the emission from  $\bh$  pairs is expected to have its peak. We shall show 
this in \S\ref{numerical} with numerical examples that take into account the
exact injection distribution of  $\bh$  pairs. Concluding,
expressions (\ref{eq7}) and (\ref{eq8}) indicate
that the synchrotron emission from  $\bh$  pairs spans over a wide range of energies 
(see also Fig.~4 in DMPR12), and may contribute to the soft $\gamma$-ray emission affecting therefore
the spectral shape of the SED.

\subsection{Proton energy loss rates}
\label{A2}
The proton energy loss rate due to the $i$ process is defined
as  $t_{\rm i}^{-1} =  -(d\gp/dt)_{\rm i} \gp^{-1}$. 
The energy loss timescale due to $\pg$ interactions
is given by (\citealt{stecker68, begelmanrudak90}-- henceforth, BSR90):
\eqb
\tpg^{-1}\left(\gp \right) =\frac{c}{2\gamma_{\rm p}^2} 
\int_{\bar{\epsilon_{\rm th}}}^{\infty} d\bar{\epsilon} \sigma_{\pg}(\bar{\epsilon}) \kappa_{\rm p}(\bar{\epsilon})\bar{\epsilon}
\int_{\bar{\epsilon}/ 2\gamma_{\rm p}}^{\infty} d\epsilon' \frac{n'(\epsilon')}{\epsilon^{' 2}},
\label{eq9}
\eqe
where bared quantities are measured in the proton's rest frame, $\bar{\epsilon}_{\rm th} = 145$~MeV, $\sigma_{\pg}$ and $\kappa_{\rm p}$
are the cross section and proton inelasticity, respectively, and $n'(\epsilon')$ is the photon number density in the comoving frame of the emission region.
Above the threshold for $\pg$ production the cross section is dominated by the $\Delta$(1232) resonance ($\sigma_{p\pi}\simeq 0.5$~mb). 
Various other resonances, albeit less important than the $\Delta(1232)$ 
resonance, shape the cross section up to $\bar{\epsilon}  \sim 1$~GeV. At higher energies though, the cross section becomes
approximately energy independent with a value $\sim 0.1$mb \citep{muecke00, beringeretal12}. 
The two-step function approximation of $\sigma_{\pg}$
presented in \cite{atoyandermer01} is an elaborate choice, 
which also takes into account the increase of the proton inelasticity as 
the interactions occur away from the threshold. However, since the exact numerical value of the proton loss
rate is not central in our case, we adopt the more crude, yet simpler, approximation
$\sigma_{\pg} \approx \sigma_0 H(\bar{\epsilon} - \bar{\epsilon}_{\rm th})$ with $ \sigma_0 \approx 1.5 \times 10^{-4}\sth$
(see also \citealt{petromast12}), and a constant inelasticity $\kappa_{\rm p}=0.2$.

The proton energy loss rate for the  $\bh$  process on an isotropic photon field was derived by 
\cite{blumenthal70} -- henceforth B70, and is given by
\eqb
\tbh^{-1}\left(\gp \right) = \frac{3}{8 \pi \gp}\sth c\af \frac{\mel}{\mpr} \int_{2}^{\infty} d\kappa \ n'\left(\frac{\kappa}{2\gp}\right)\frac{\phi(\kappa)}{\kappa^2},
\label{eq10}
\eqe
where $\af$ is the fine structure constant, $\kappa = 2\gp \epsilon'/\mel c^2$, 
and  $\phi(\kappa)$  is a function defined by a double integral (see eq.~(3.12) in \citealt{chodorowski92}; henceforth CZS92).
CZS92 derived analytical approximate expressions for  $\phi(\kappa)$, yet the analytical calculation
of the integral in eq.~(\ref{eq10}) is cumbersome even for the case of a power-law photon distribution (see appendix).

For BL~Lac objects, the main contribution to $n'(\epsilon')$ comes 
from the synchrotron radiation of primary electrons.
We express the photon number density in terms of observable quantities, such as the bolometric synchrotron luminosity 
$L_{\rm syn}$. The energy density of synchrotron photons in the comoving frame is given by
\eqb
u'_{\rm syn} \approx \frac{3 L_{\rm syn}}{4 \pi \delta^4 \rb^2 c},
\label{eq11}
\eqe
where $\rb$ is the comoving size of the emission region. In the limit of $\theta \lesssim 1/\Gamma$ and $\Gamma \gg 1$, we find
$\delta \approx \Gamma$. From this point on, we will use interchangeably $\delta$ and $\Gamma$. 
 The low-energy spectrum of blazars, especially that of LBLs\footnote{This is also true 
for the SED of the radio galaxy Centaurus A (e.g. \citealt{petrolefa14}).}, can often be described by a steep power-law for energies above 
its peak in $\epsilon L(\epsilon)$ units. For indicative examples, see Fig.~2 in \cite{ghisellini01} and Fig.~5 in \cite{boettcherreimer13}. For this reason,
we derive explicit expressions of $\tpg^{-1}$ and $\tbh^{-1}$ for a power-law synchrotron photon distribution.
As a first step though, and for completeness reasons, we examine the case of a monoenergetic photon distribution. This 
can be considered as a zero order approximation
for a narrow (in energy) low-energy hump, and faciliates the derivation of simple analytic relations.

\subsubsection{Monoenergetic photon distribution}
We assume that the synchrotron photon field is monoenergetic
with energy $\es$.  The differential photon number density in the comoving frame is then written as
\eqb
n'(\epsilon') = n'_0\epsilon' \delta(\epsilon'-\esco),
\label{eq12}
\eqe
with $n'_0=u'_{\rm syn}/ \epsilon_{\rm s}^{'2}$.  We note that the rectangular approximation of the $\pg$ cross section, which
is adequate for the case of power-law photon distributions (e.g. \citealt{murasedermer14}), 
 is not appropriate in this case, since it understimates the cooling of protons with energies much above the threshold.
 Using eqs.~(\ref{eq9}) and (\ref{eq12}) we find the energy loss rate to be
\eqb
\label{eq15a}
\tpg^{-1} \left(\xi_{\pg} \right) \approx  7 \times 10^{-5} \ s^{-1}\left(1-\frac{1}{\xi_{\pg}^2} \right)
\frac{L_{\rm syn, 45}}{r^2_{\rm b, 15} \delta^3 \nu_{\rm s, 16}  \left(1+z \right)}, \ \xi_{\pg} >1,
\eqe
 where $\xi_{\pg}=2\gp/\gamma_{\rm p, \pg}^{\rm (th)}$. 
The dimensionless energy loss rate is defined as 
$\tpg^{-1}/\tcr^{-1}$, where $\tcr=\rb/c$ and is also an efficiency measure of the process. This is written as
\eqb
\label{eq15b}
\fpg (\xi_{\pg})  \approx 2.2 \left(1-\frac{1}{\xi_{\pg}^2} \right)
\frac{L_{\rm syn, 45}}{r_{\rm b, 15} \delta^3 \nu_{\rm s, 16}  \left(1+z \right)}, \ \xi_{\pg} >1.
\eqe
The loss rate becomes constant for $\gp$ slightly above the threshold,  which is in rough agreement with numerical
calculations where the full cross section and the energy-dependent inelasticity are used.

Subtitution of eq.~(\ref{eq12}) into eq.~(\ref{eq10}) leads to
\eqb
\label{eq16}
\tbh^{-1}(\gp) & = &  \frac{3}{8 \pi}\sth c \af \frac{\mel}{\mpr} n'_0 \mel c^2 \frac{\phi\left(2\gp \xs \right)}{2\gp^2 \xs}, \ \gp > \frac{1}{\xs} \\
\tbh^{-1}(\xi_{\rm  \bh }) & \simeq  &  2 \times 10^{-6}\  s^{-1}\ g(\xi_{\rm   \bh }) \frac{L_{\rm syn, 45}}{r^2_{\rm b, 15} 
\delta^3  \nu_{\rm s, 16} \left(1+z \right) }, \xi_{\rm \bh } > 1,
\label{eq17}
\eqe
where $\xi_{\rm  \bh } = 2 \gp/\gamma_{\rm p,  \bh }^{\rm (th)}$ with $\gamma_{\rm p,  \bh }^{\rm (th)}$ given by eq.~(\ref{eq4})
and $g(\xi) = \phi (\xi)/\xi^2$, which has its maximum $g_{\max} \sim 1$ at $\xi \simeq 47$ (see Fig.~2 in CZS92). 
Similarly to the $\pg$ efficiency, for the $\bh$  process we find
\eqb
\label{eq18}
\fbh(\xi_{\rm  \bh }) \approx 0.06 g(\xi_{\rm   \bh }) \frac{L_{\rm syn, 45}}{r_{\rm b, 15} 
\delta^2 \nu_{\rm s, 16}  \left(1+z \right) }, \xi_{\rm   \bh } > 1.
\eqe
As the dependence on the parameters related to blazar emission is the same in eqs.~(\ref{eq15b}) and (\ref{eq18}),
we find that $\fbh \ll \fpg$ for $\gp \ge \gamma_{\rm p, \pg}^{\rm (th)}$ in the case of monoenergetic photons.
 
\subsubsection{Power-law photon distribution}
\begin{figure*}
\includegraphics[width=0.48\textwidth]{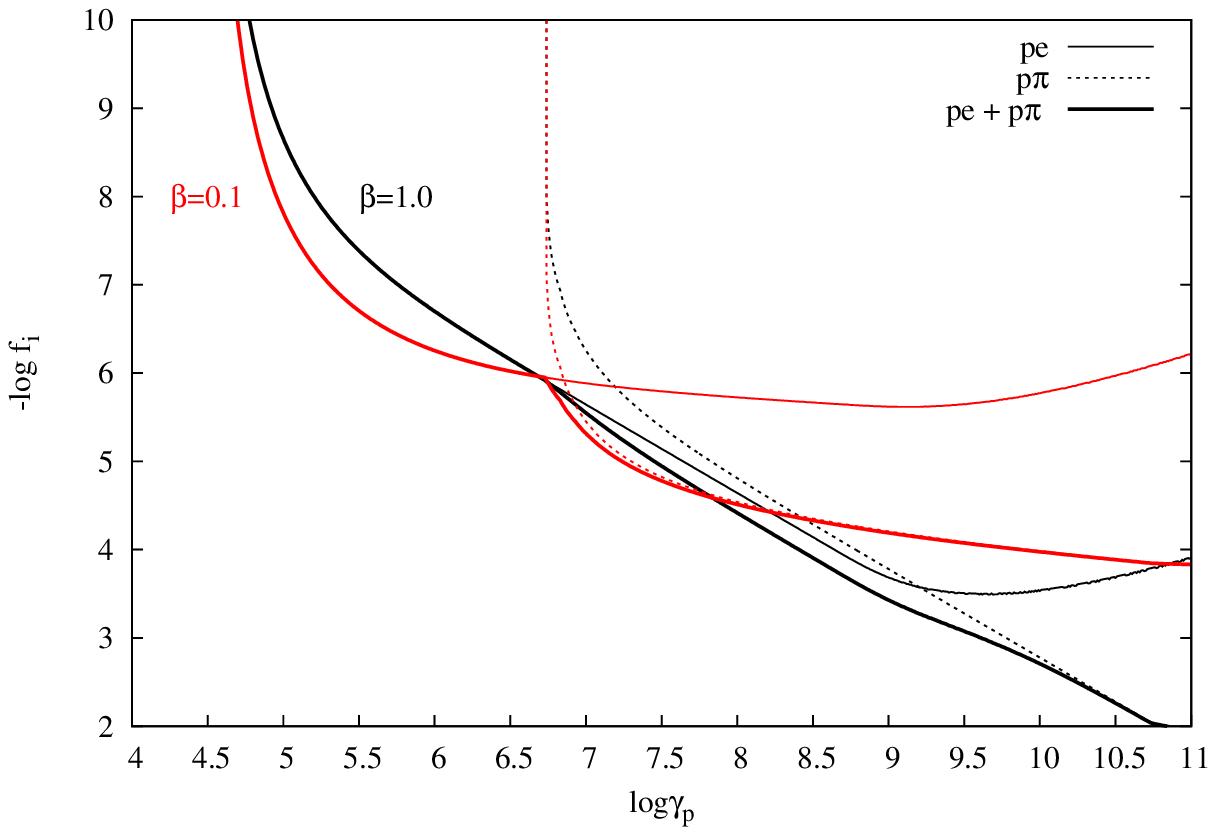} 
\includegraphics[width=0.48\textwidth]{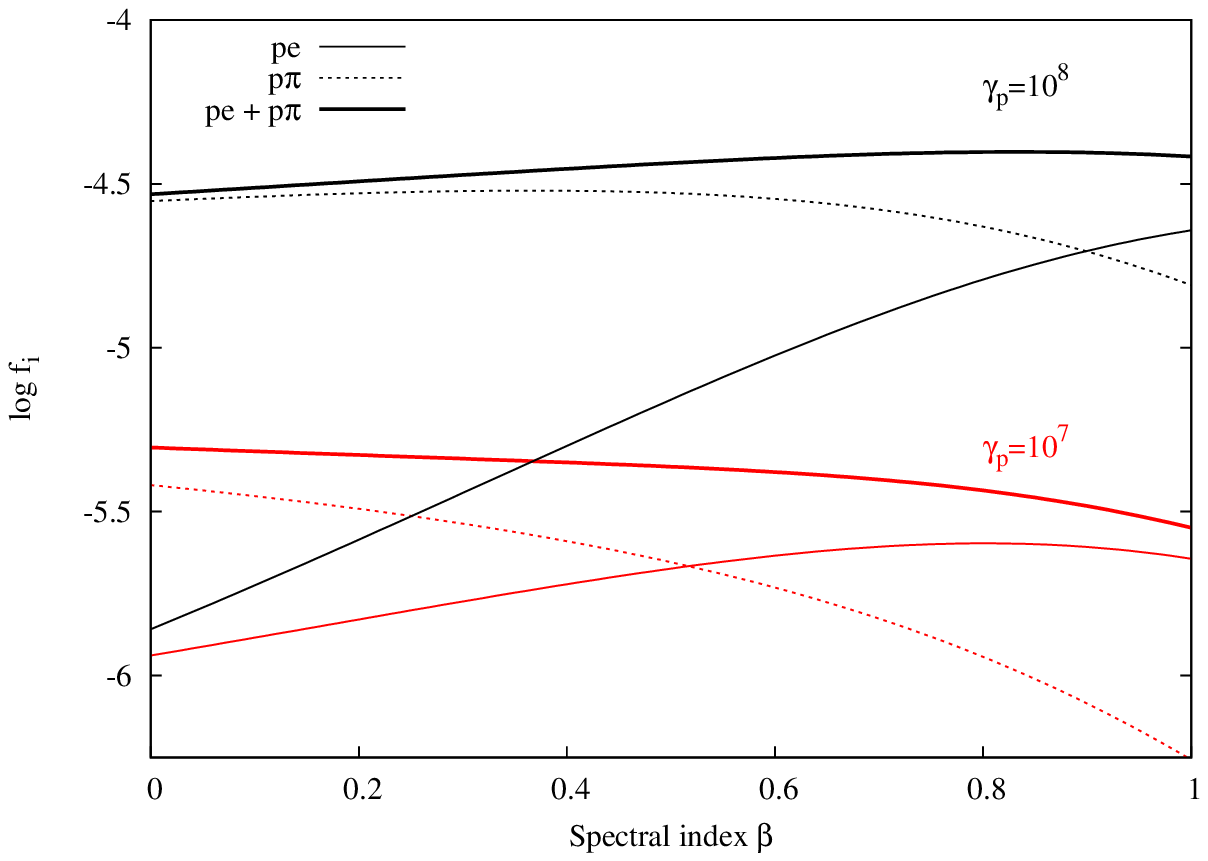}
\caption{Left panel: Logarithmic plot of the dimensionless loss timescales $\fbh^{-1}$ (thin solid lines), $\fpg^{-1}$ (dotted lines), and $\fpg^{-1}+\fbh^{-1}$
(thick solid lines) as a function of $\gp$ for two spectral indices $\beta$ of the photon distribution marked on the plot.
Right panel: Plot of the dimensionless loss rates $\fbh$ (thin solid lines), $\fpg$ (dashed lines),
and $\fpg+\fbh$ (thick solid lines) as a function $\beta$ for $\gp=10^{7}$ (red lines) and $\gp=10^8$ (black lines).
Other parameters used in both panels: $L_{\rm syn}=10^{45}$~erg/s, $\rb=10^{15}$~cm, $\nu_{\rm s}=10^{17}$~Hz, $\emin=10^{-4}\emax$,
$\es=\emax$ and $\delta=30$.}
\label{fig1}
\end{figure*}
\begin{figure*}
\includegraphics[width=0.49\textwidth]{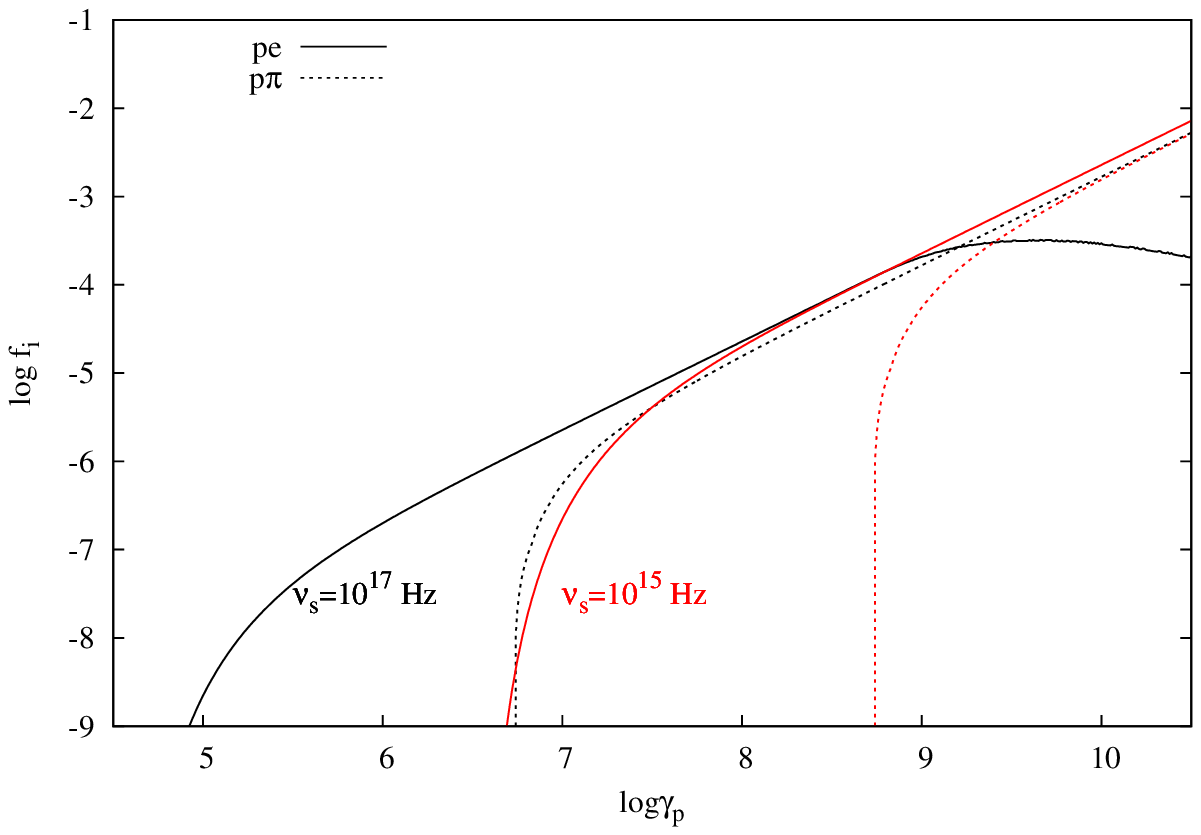} 
\includegraphics[width=0.49\textwidth]{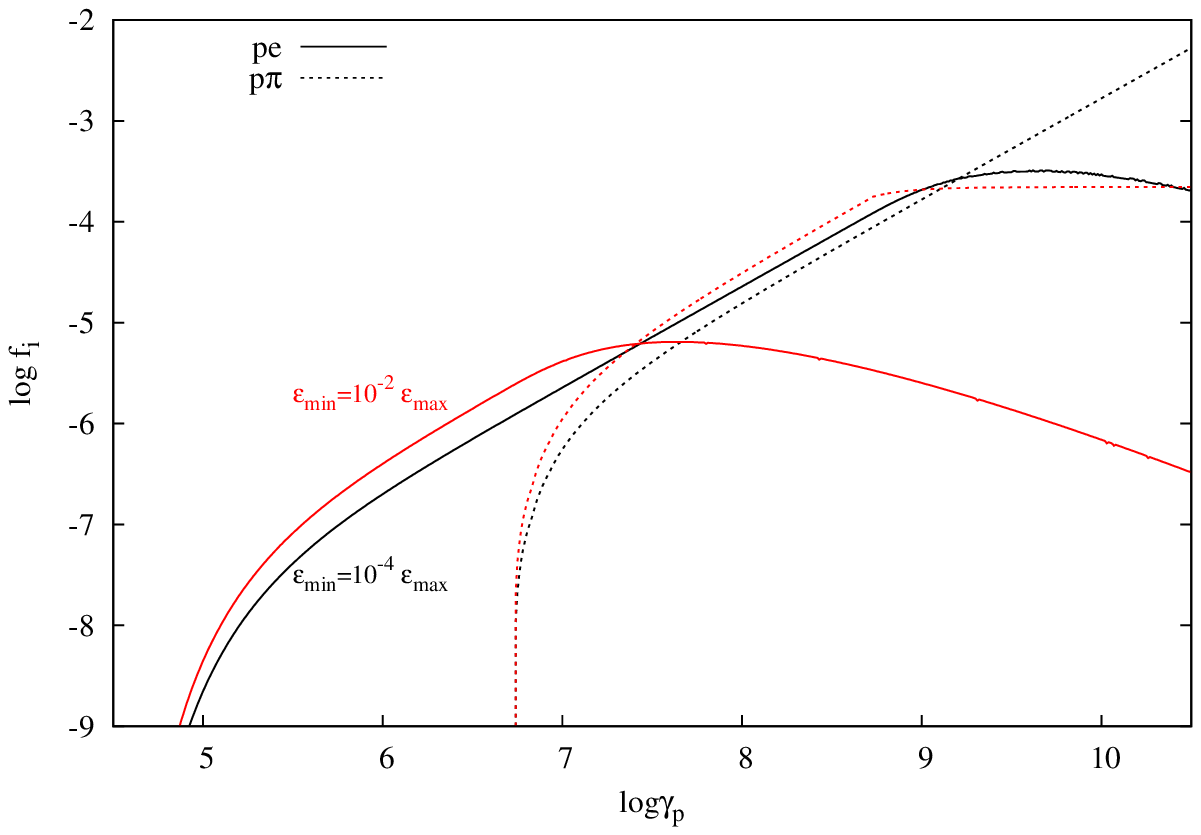}
\caption{Logarithmic plots of the dimensionless loss rates $\fbh$ (solid lines) and $\fpg$ (dotted lines)
as a function of $\gp$. 
Left panel: Dependence on the high cutoff energy of the synchrotron spectrum 
for $\emin=10^{-4} \emax$. Black and red lines
correspond to $\nu_{\rm s}=10^{17}$~Hz and $\nu_{\rm s}=10^{15}$~Hz, respectively.
Right panel: Dependence on the low cutoff energy of the synchrotron spectrum for $\nu_{\rm s}=10^{17}$~Hz.
Black and red lines 
correspond to $\emin/\es=10^{-4}$ and $10^{-2}$, respectively.
Other parameters used are: $L_{\rm syn}=10^{45}$~erg/s, $\rb=10^{15}$~cm,  $\delta=30$, $\beta=1$ and $\es \equiv \emax$.}
\label{fig2}
\end{figure*}
We assume that the differential
luminosity scales as $L_{\rm syn}(\epsilon) \propto \epsilon^{-\beta}$  with
$\beta >0$ for $\emin< \epsilon < \emax$,
and normalize $L_{\rm syn}(\epsilon)$ with respect to the 
 peak energy $\es= h \nu_{\rm s}$ (see point (i) in \S\ref{A1}), which 
is set equal to $\emax$ for $\beta \le 1$ or $\emin$ otherwise.
Given that 
\eqb
\label{eq19}
L_{\rm syn}(\epsilon) = \frac{L_0}{\es}\left(\frac{\epsilon}{\es} \right)^{-\beta}
\eqe
with the normalization $L_0$ being
\eqb
\label{eq20}
\frac{L_0(\beta, \esco)}{L_{\rm syn}}  =  \left\{ \begin{array} {cc}
                           (1-\beta) \left(\frac{\es}{\emax} \right)^{-\beta+1} \left[1-\left(\frac{\emin}{\emax}\right)^{-\beta +1}\right]^{-1}, & \beta \neq 1 \\
                              & \\
                           1/\ln\left(\frac{\emax}{\emin} \right), &  \beta = 1,
                           \end{array}
                           \right.
\eqe
we may write the differential photon number density in the comoving frame  as
\eqb
n'(\epsilon') = n'_0 \left(\frac{\epsilon'}{\esco}\right)^{-\beta-1}.
\label{eq21}
\eqe
The respective normalization is 
\eqb
n'_0 = \frac{3 L_0(\beta, \esco)}{4\pi c \delta^4 \rb^2 \epsilon_{\rm s}^{' 2}}.
\label{eq22}
\eqe
By inserting expression (\ref{eq21}) into eq.~(\ref{eq9}) 
we calculate $\tpg^{-1}$, which
is now written as
\eqb
\tpg^{-1}(\gp)=  \frac{3 L_{\rm syn} \lambda(\beta, \es) \kappa_{\rm p} \sigma_0}{2 \pi \rb^2 \delta^3 \es (1+z)} 
 \left \{\begin{array}{cc}       
     \left(\frac{2 \gp \esco}{\overline{\epsilon}_{\rm th}} \right)^{\beta} - \left(\frac{\esco}{\epsilon'_{\max}} \right)^{\beta}, & 
     \gp < \frac{\bar{\epsilon}_{\rm th}}{2\epsilon'_{\min}}  \\
     & \\
    \left(\frac{\es}{\emin} \right)^{\beta} & 
     \gp > \frac{\bar{\epsilon}_{\rm th}}{2\epsilon'_{\min}},  \\
     \end{array}  
     \right.
\label{eq23}     
\eqe
 where we neglected the contribution of the upper cutoff when performing the integral over the photon energies $\epsilon'$.
We also define $\lambda$ as
\eqb
\lambda(\beta, \es)= \frac{L_0(\beta, \esco)}{\beta (\beta+2)L_{\rm syn}}, \ \beta \ne 0, -2.
\label{eq24}
\eqe
 We note that the first branch of eq.~(\ref{eq23}) coincides 
with eq.~(5) in \cite{mannheim91} after making the replacements $2 n'_0 (\esco)^{\beta+1} \rightarrow m_0$, $\beta \rightarrow \alpha_{\rm t}$,
$\sigma_0 \kappa_{\rm p} \rightarrow \langle \kappa \sigma_{\rm p\gamma \rightarrow \Delta}\rangle$, and for
$\gamma_{\rm p} \gtrsim \gamma_{\rm p, \pg}^{\rm (th)}\es/2 \epsilon_{\max}$.
It agrees also with the expression derived by eq.~(4) in BSR90 for 
$\sigma^{\pi}(x')K^{\pi}(x') \approx\overline{\sigma^{\pi}(x')K^{\pi}(x')}$, where the latter corresponds to the product 
$\sigma_0 \kappa_{\rm p}$ in this work. 
Since BSR90 assumed that the power-law photon distribution extends to very low energies, as to establish 
$\gp \epsilon'_{\min} \lesssim \bar{\epsilon}_{\rm th}$,
the second branch of eq.~(\ref{eq23}) was not relevant in their analysis.

In terms of the variable $\xi_{\pg}$ the photopion efficiency is written
as
\eqb
\fpg(\xi_{\pg}) \simeq 4.4 \frac{L_{\rm syn, 45}\lambda(\beta, \es)}{r_{\rm b, 15} \delta^3 \nu_{\rm s, 16}(1+z)}
\left \{\begin{array}{cc}       
     \xi_{\pg}^{\beta}, & 
     \xi_{\pg} <\frac{\es}{\emin}  \\
     & \\
    \left(\frac{\es}{\emin} \right)^{\beta}, & 
   \xi_{\pg} >  \frac{\es}{\emin}  \\
     \end{array}  
     \right.
\label{eq25}
\eqe
We neglected the term $\left(\esco/\epsilon'_{\max}\right)^{\beta}$ from the first branch of eq.~(\ref{eq25})
since it is always less than unity and we are interested in $\xi_{\pg} \gtrsim 1$.

We found that a similar expression for the photopair production 
can also be derived, if the function
$\phi(\kappa)/\kappa^2$ that appears in the integral
of eq.~(\ref{eq10}) is approximated by a bi-Gaussian function with respect
to $\ln\kappa$ (see appendix for more details). The dimensionless loss rate may be written as:
\eqb
\fbh(\xi_{\rm  \bh }) \simeq 0.06 \ \frac{L_{\rm syn, 45} 
\beta(\beta + 2)\lambda(\beta, \es)}{r_{\rm b, 15} \delta^3 \nu_{\rm s, 16}(1+z)} \xi_{\rm  \bh }^{\beta} I(\gp,\beta),
\label{eq25b}
\eqe
where $I(\gp,\beta)$ is a function that may be expressed in terms
of error functions.  Although general use of eq.~(\ref{eq25b}) may be cumbersome because of 
the presence of error functions, it can be used for having a first estimate
of the  $\bh$ loss rate in the case of a power-law photon distribution. It also faciliates 
the comparison between the loss rates for the two channels of photohadronic interactions, 
since the expressions of eqs.~(\ref{eq25}) and (\ref{eq25b}) 
are similar, apart from the factor
 $I(\gp,\beta)$.  The ratio of the loss rates for a fixed proton energy 
is $\fpg(\gp)/\fbh(\gp) \approx 73 [(\beta+2)\beta]^{-1} 
\left(\gamma_{\rm p,   \bh }^{\rm (th)}/ \gamma_{\rm p, \pg}^{\rm (th)}\right)^{\beta} I(\gp, \beta)^{-1}$.
The ratio of the two rates becomes of order unity for 
$\beta \sim 1$ (see also BRS90), where we also used the fact that $I(\gp, 1) \sim 0.1$ for a wide range of $\gp$ values
(see Fig.~A3 in appendix).

In what follows, we numerically integrate eqs.~(\ref{eq9}) and (\ref{eq10}) for a power-law photon distribution and
compare the derived $\fpg$ and $\fbh$ for different blazar parameters\footnote{We first tested the numerical integration scheme
used for the calculation of eqs.~(\ref{eq9}) and (\ref{eq10}) by applying the same parameters used for Fig.~2 in \cite{begelmanrudak90}, i.e.
$\beta=\alpha=1$, $u_{\rm rad} \simeq 10^{4}$~erg/cm$^3$, $x_{\max}=\epsilon'_{\max}/\mel c^2 =0.5$ and $x_{\min}=\epsilon'_{\min}/\mel c^2 =10^{-7}$.}. 
The results are summarized in
Figs.~\ref{fig1} and \ref{fig2}.

In general, we find that
\begin{itemize}
 \item below the threshold energy for photopion production, proton losses are dominated by photopair production as expected 
 (e.g. \citealt{stanevetal00}, DMPR12).
 \item in general, $\pg$ losses dominate above the respective energy threshold but there are parameters leading to similar, 
 at least within the same order of magnitude,
 energy loss rates. For example, we find 
 $\eta = \fbh / \fpg \sim 0.1-1$, for  $\gp \gtrsim 
 \gamma_{\rm p, \pg}^{\rm (th)}$ and $\gp \gg \gamma_{\rm p,   \bh}^{\rm (th)}$ (left panel in Figs.~\ref{fig1} and \ref{fig2}).
 \item softer synchrotron spectra, namely larger $\beta$, favour $\bh$ pair production, as shown in the right panel of Fig.~\ref{fig1}.
 \item wider synchrotron spectra, i.e. larger 
 value of the ratio $\es/\emin$, favours also the  $\bh$  process (right panel in Fig.~\ref{fig2}).
 \item higher peak frequencies push the threshold  Lorentz factors to lower values, as eqs.~(\ref{eq3}) and (\ref{eq4}) show. Thus, for
 a fixed proton energy much above the threshold for  $\bh$ pair production , higher $\nu_{\rm s}$ translates to lower $\fbh$; 
 this can be seen in the left panel of Fig.~\ref{fig2}.
 \item smaller values of the Doppler factor result in: an increase of both $\fbh$ and $\fpg$ (see e.g. eqs.~(\ref{eq25}) and (\ref{eq25b})) 
 and a decrease of the respective threshold Lorentz factors (see eqs.~(\ref{eq3}) and (\ref{eq4})).
\end{itemize}

\subsection{Luminosity estimates}
\label{remarks}
If $L_{\pg}$ and $L_{\rm \bh }$ denote the bolometric synchrotron luminosities from  pairs
produced by the photopion and photopair processes, respectively, 
and under the assumption of  efficient cooling of pairs, we may write $L_{\rm \bh} \approx \fbh L_{\rm  p}$
and $L_{\pg} \approx (\fpg/8) L_{\rm p}$,  where $L_{\rm p}$ is the proton luminosity.
The latter is derived under the assumptions that approximately half of the produced pions 
are neutral, thus not contributing to the injection of pairs, and that the produced electron/positron 
acquires $\sim 1/4$ of the proton's energy
in each $\pg$ collision. The luminosity ratio of the two components is then given by
$L_{\rm   \bh }/L_{\pg} \approx 8 \eta$.
If we combine this estimate with the fact that the typical energy of synchrotron photons emitted
by the  $\bh$  pairs for $\gp \gg \gamma_{\rm p,   \bh }^{\rm (th)}$ 
falls in the hard X-ray/soft $\gamma$-ray energy range  (see \S\ref{A1}),
we expect 
a third bump in the SED with luminosity $\sim 8\eta$ of the $\gamma$-ray luminosity.
Thus, for parameters that lead to $\eta \sim 0.1-1$ the additional component has the same order of magnitude luminosity with the $\gamma$-rays.
Interestingly, even if $\eta \ll 1$, we expect that the  $\bh$ emission can still  bridge the two main humps of the SED, since
the synchrotron emission from  $\bh$ pairs does not, in general, overlap
with the one from $\pg$ produced pairs (see e.g.~Fig.~\ref{fig0}).

Going one step further, we may relate $L_{\rm  \bh }$ with the luminosity emitted in neutrinos, which 
are a by-product of $\pg$ interactions.
Let us first estimate what is the typical neutrino energy, which in $\mel c^2$ units and
in the comoving frame is written as
\eqb
x_{\nu}  \approx  \frac{1}{4}\kappa_{\rm p} \gamma_{\rm p} \frac{\mpr}{\mel}.
\label{eq8a}
\eqe
For protons with $\gp = \gamma_{\rm p, \pg}^{\rm (th)}$ this translates to an observed energy 
\eqb
\epsilon_{\nu}^{\rm (th)} \simeq 0.2 \ {\rm PeV} \ \delta^2 (1+z)^{-2} \nu_{\rm s, 16}^{-1},
\label{eq8b}
\eqe
where we used eq.~(\ref{eq3}) and the superscript `th' is used as reminder for the parent proton's energy. 
The ratio $\mathcal{R}_{\nu, \gamma}$ of the typical neutrino ($x_{\nu}$)
and synchrotron photon energies from $\pg$ pairs ($x_{\rm s}^{\pg}$) is given by
\eqb
\mathcal{R}_{\nu,\gamma} \equiv \frac{x_{\nu}}{x_{\rm s}^{\pg}}  = 4 \times 10^3 \delta^2 (1+z)^{-2} \left( \nu_{\rm s, 16} \nu_{\gamma, 25} \right)^{-1},
\label{eq8c}
\eqe
where we used 
\eqb
x_{\rm s}^{\pg}  \approx  b \left( \frac{1}{4}\kappa_{\rm p} \gamma_{\rm p} \frac{\mpr}{\mel}  \right)^2,
\label{eq8d}
\eqe
as well as eqs.~(\ref{eq3}),(\ref{eq5}), (\ref{eq8a}) and $b=B/\Bcr$.
In this context, the ratio $\mathcal{R}_{\nu, \gamma}$ is an estimate of the
energy separation of the $\gamma$-ray and neutrino components. It increases
quadratically with the Doppler factor, while it decreases for higher synchrotron peak frequencies. 
Thus, for a fixed Doppler factor,
the separation of the neutrino and $\gamma$-ray components decreases as we move from LBLs to HBLs. The same applies
to the neutrino energy, which
moves to lower values as $\nu_{\rm s}$ increases (see also \citealt{murasedermer14}).
If $L_{\nu}$ is the total luminosity in electron
and muon neutrinos, we find that  $L_{\nu} \approx 3 L_{\pg}$ and  $L_{\rm  \bh } \approx (8/3)\eta L_{\nu}$.
In our analytical estimations, the $\gamma$-ray emission results
from $\pg$ pairs that have as parent particles, protons with energy close to the threshold energy
(see point (iv) in \S\ref{A1}). Thus, we find  $L_{\rm  \bh } \sim L_{\pg} \sim L_{\nu}$ for $\eta \sim 0.1-1$, whereas
if $\eta \ll 1$ we expect $L_{\rm  \bh } \ll L_{\pg} \sim L_{\nu}$.

In the following section we will discuss the above predictions through
detailed numerical examples using parameters relevant to blazar emission.

\section{Numerical approach}
\label{numerical}
\subsection{Numerical code}
The results presented in this section are obtained
using a numerical code developed for solving systems
of coupled integrodifferential equations.
For the physical scenario we investigate, the system 
consists of five equations, one for each stable particle species, namely
protons, electrons, photons, neutrons and neutrinos. 
The various rates are written in such a way as to ensure self-consistency, i.e. the amount of
energy lost by one species in a particular process is equal to that emitted (or injected)
by another. That way one can keep the logistics of the system in the sense that
at each instant the amount of energy entering the source through the injection 
of protons and primary electrons should equal the amount of energy escaping from it in the form of 
photons, neutrons and neutrinos; to this one has to include the energy carried away because of the 
electron and proton physical escape from the source. 

These advantages of the kinetic equation approach are also combined with the detailed modeling
of photohadronic interactions using results from Monte Carlo simulations. In particular, for 
$\bh$ pair production the Monte Carlo results by \cite{protheroe96} were used  (see also \citealt{mastetal05}). Photopion interactions
were incorporated in the time-dependent code by using the results of the Monte Carlo event generator SOPHIA \citep{muecke00}.
More details about the rates of various processes can be found
in DMPR12, while description of 
additional improvements, e.g. inclusion of pion, muon and kaon synchrotron cooling, are presented
in \cite{petrogianniosdim14}.

The free parameters of the model, which are used as an input to the numerical code 
are summarized below:
\begin{enumerate}
\item the radius $\rb$  and magnetic field $B$ of the emission region; 
\item its Doppler factor, $\delta$;
\item the injected luminosities of
protons and primary electrons, which are expressed in terms 
of compactnesses:
\eqb
\ell_{\rm i}^{\rm inj}={{L_{\rm i} \sth}\over{4\pi \rb \delta^4 m_{\rm i} c^3}},
\label{lpinj}
\eqe
where $i$ denotes protons or electrons;
\item the physical escape time
for both particles, which is assumed to be the same and equal to the crossing time of the source, i.e.
$t_{\rm p, esc}=t_{\rm e, esc}=\tcr$;
\item the maximum and minimum Lorentz factors of the injected protons and primary electrons, 
$\gamma_{\rm i, \max}$ and $\gamma_{\rm i, \min}$ respectively; and
\item the power-law indices $s_{\rm p}$ and $s_{\rm e}$ of injected
protons and primary electrons, respectively.
\end{enumerate}

\subsection{Results}
We present indicative examples of multiwavelength (MW) photon and neutrino spectra calculated within our scenario
and interpret them using the insight gained from the analysis of \S\ref{A1} and \S\ref{A2}. Application of the model
to specific BL~Lac sources will be presented elsewhere (Petropoulou et al. 2014, in preparation).
We present two baseline models 
with parameter values that differ significantly, e.g. $B=0.1$~G (Model A)
and $B=10$~G (Model B), in order to demonstrate that the appearance of the
$\bh$  component in the SED  is not just the result of a very specific parameter choice but a  generic feature 
of leptohadronic emission models, which is  often overlooked. 
All model parameters are summarized in Table~\ref{tab0}. 
For each baseline model,  we then create a template of variants by changing only one parameter each time, in order 
to understand their impact on the SED.
In all examples, we used a fiducial redshift $z=0.14$ and corrected the high-energy part of the spectra for 
photon-photon absorption on the extragalactic background light (EBL) using
the EBL model of \cite{franceschini08}.
\subsubsection{Photon emission}
\label{photons}
\begin{table}
\centering
 \caption{Input parameters for the baseline models discussed in text.}
 \begin{tabular}{ccc}
  \hline\hline
  Parameter & Model~A & Model~B \\
  \hline\hline
  B (G) & 0.1 & 10  \\
  $\rb$ (cm) &  $3\times 10^{16}$ & $3\times 10^{15}$  \\
  $\delta$ & 30 & 15  \\
  \hline
  $\gamma_{\rm e, \min}$ & 1 & $3\times 10^2$   \\
  $\gamma_{\rm e, \max}$ & $3\times 10^5$  & $3\times 10^6$  \\
   $s_{\rm e} $ & 2.0 & 2.5  \\
  $\leinj$ & $1.2\times 10^{-6}$ & $2 \times 10^{-3}$  \\
  \hline
  $\gamma_{\rm p, \min}$ & 1 & 1 \\
  $\gamma_{\rm p, \max}$ & $1.2 \times 10^7$ & $6.3\times 10^6$  \\
  $s_{\rm p}$ & 2.0 & 2.0 \\
    $\lpinj$  & $10^{-3}$ & $1.2\times 10^{-2}$ \\
  \hline
 \end{tabular}
\label{tab0}
 \end{table}
 \begin{figure}
\includegraphics[width=0.45\textwidth]{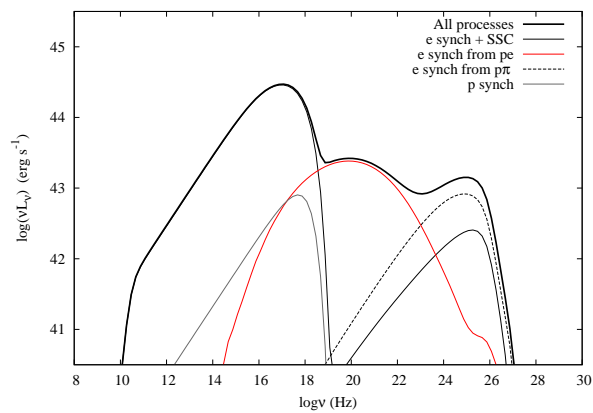} 
\caption{SED of the baseline Model A (thick solid line). The primary leptonic synchrotron and SSC components
are plotted with thin solid line. Synchrotron emission from $\pg$ pairs is shown with 
black dashed lines, while the $\bh$  synchrotron spectrum is plotted with red solid line. 
Proton synchrotron radiation is overplotted with a grey line.}
\label{fig6}
\end{figure}
\begin{figure}
\includegraphics[width=0.45\textwidth]{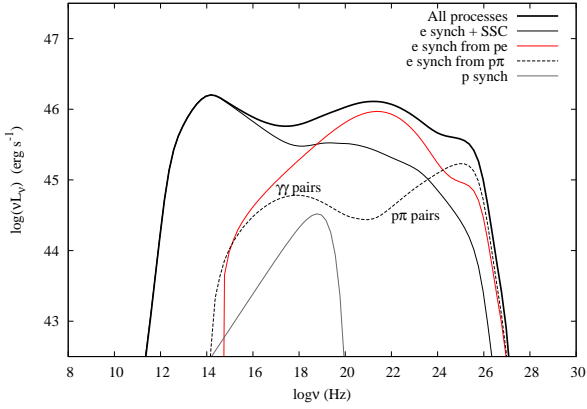} 
\caption{SED of the baseline Model B (thick solid line).
The primary leptonic synchrotron and SSC components
are plotted with thin solid line. 
The synchrotron spectrum of $\bh$ pairs is plotted with red solid line, while
synchrotron emission from secondary pairs produced through $\pg$ interactions
and photon-photon absorption of VHE $\gamma$-rays is shown with 
black dashed lines. Proton synchrotron radiation is overplotted with a grey line.
}
\label{fig4}
\end{figure}
\begin{figure}
\includegraphics[width=0.45\textwidth]{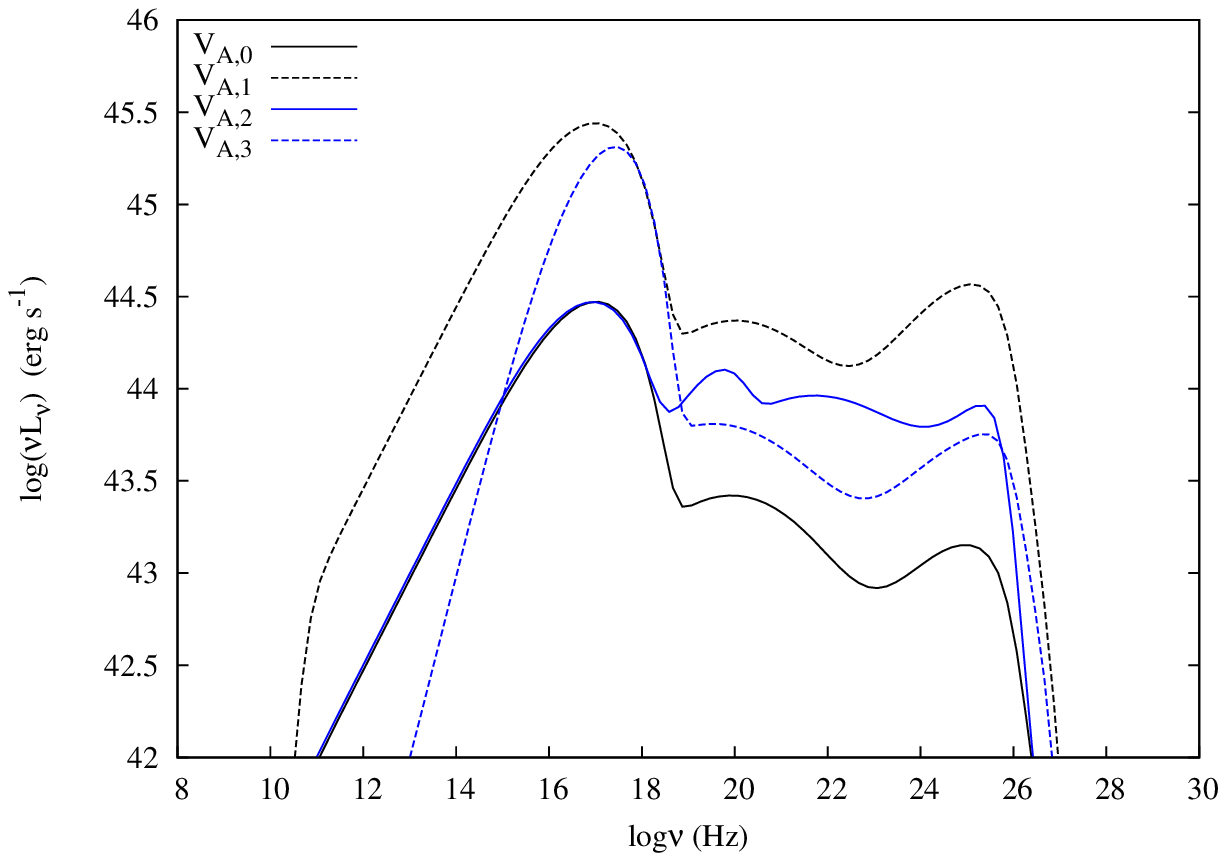} 
\caption{MW photon spectra obtained for the variants of the baseline Model~A discussed in text.}
\label{fig7}
\end{figure}

\begin{figure}
\includegraphics[width=0.45\textwidth]{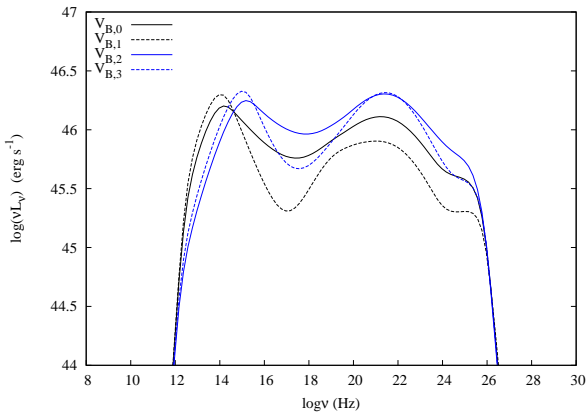} 
\caption{MW photon spectra obtained for the variants of the baseline Model~B discussed in text.
}
\label{fig5}
\end{figure}
The MW photon spectra obtained for the baseline Model A are shown in Fig.~\ref{fig6}.
The total emission when all processes are taken into account is plotted with a thick solid line.
The derived synchrotron luminosity and peak frequency in this model are 
respectively $L_{\rm syn} \simeq 7 \times 10^{44}$~erg/s and $\nu_{\rm s}\simeq 10^{17}$~Hz,
which are more relevant to HBLs (e.g. \citealt{ghisellinitavecchio08}). Application of 
the model to a particular
HBL source, namely the prototype blazar Mrk~421, can be found in \cite{mastetal13} and \cite{dimitrakoudis14}.
The primary electron synchrotron and SSC emission are plotted with thin solid lines. The synchrotron emission of  
$\bh$  and $\pg$ pairs is shown with red solid and black dashed lines,
respectively. Finally, the subdominant proton synchrotron component is plotted
with a grey line. Although proton losses because of  $\bh$ pair production and photopion production are
similar, the injection spectra of secondaries are different, thus resulting in very different 
emission signatures (see e.g. Figs.~7 and 4 in \citealt{kelneraharonian08} and DMPR12, respectively). Of particular interest is the broad
synchrotron component of $\bh$  pairs, which features photopair production
as a physical process for the production of  wide curved spectra outside the usual SSC framework.

We present the photon spectra for the baseline Model B in Fig.~\ref{fig4}, where the 
different types of lines have the same meaning
as in Fig.~\ref{fig6}. The luminosity and peak frequency of the low-energy hump in this case are 
$L_{\rm syn}\simeq 3\times 10^{46}$~erg/s and $\nu_{\rm s}\simeq 10^{14}$~Hz, respectively, making
Model B relevant to LBL emission. Similar to Model~A, the 
contribution of the photopair synchrotron emission to the SED
is dominant for a wide range of frequencies ($\sim 10^{18}-10^{24}$~Hz).
Synchrotron emission from $\pg$ pairs, on the other hand, has a  harder spectrum than the 
emission from  $\bh$ pairs, and peaks at $\sim 10^{25}$~Hz.
An additional feature of Model~B is a component that peaks at $\sim 10^{16}$~Hz and
is explained as synchrotron radiation of pairs produced through (internal) photon-photon absorption.
In particular, the absorption of VHE $\gamma$-rays from $\pi^0$ decay
initiates an EM cascade. This terminates whenever the synchrotron photons emitted by the pairs
do not satisfy anymore the threshold condition for absorption on the photons of the low-energy component of the SED.
For parameter values resulting 
in even higher optical depths for $\gamma$-ray absorption, the emission from the EM cascade
dominates over the other components (see e.g.~\citealt{mannheimbiermann92, petroarfani14}).

Summarizing, we showed that the emission from  $\bh$  pairs 
can fill, to an extend, the gap between the
low-and high-energy humps of the SED (Fig.~\ref{fig6})
or be the dominant photohadronic emission component in the hard X-ray/soft $\gamma$-ray regime (Fig.~\ref{fig4}).
\subsubsection*{Model~A variants}
We investigate next three variants of the baseline Model~A listed below:
\begin{itemize}
\item $V_{\rm A, 0}$ for $\leinj=10^{-5.9}, \gpmx=1.2\times10^7$ and $s_{\rm e}=2$
 \item $V_{\rm A, 1}$ for $\leinj=10^{-4.9}$
 \item $V_{\rm A, 2}$ for $\gpmx=1.2 \times 10^8$ 
 \item $V_{\rm A, 3}$ for $s_{\rm e}=1$
\end{itemize}
with $V_{\rm A,0}$ corresponding  to the baseline model.
The respective photon spectra are presented in Fig.~\ref{fig7}.
The effect that a higher $\leinj$ has on the SED is straightforward, since
it is translated to a higher photon number density in the source. When compared to the $V_{\rm A,0}$ case,
the photohadronic components of the variant $V_{\rm A, 1}$ have approximately $10$ times
higher luminosity, i.e. they depend on $\leinj$ in a linear manner. This result should be compared to 
the one presented in DMPR12 (see Fig.~7 therein), where the photon target field for photohadronic interactions
was the proton synchrotron radiation itself and the dependence on $\lpinj$ was quadratic.

Example $V_{\rm A, 2}$ demonstrates the effect of a higher $\gpmx$, which now becomes
$\sim 10 \gamma_{\rm p, \pg}^{\rm (th)}$, where $\gamma_{\rm p, \pg}^{\rm (th)}\simeq 10^7$ for $\delta=30$ and $\nu_{\rm s}\simeq 10^{17}$~Hz 
(see eq.~(\ref{eq3})).
The $\pg$ component peaks at approximately the same frequency as for the $V_{A,0}$ case, 
although the estimated shift according to eq.~(\ref{eq8d}) is two orders of magnitude. 
The reason for this discrepancy is photon-photon absorption on the EBL, which attenuates
the high energy part of the spectrum for the fiducial redhsift $z=0.14$. This
is also indicated by an abrupt steepening of the high-energy part of the spectrum.
Because the number density of protons depends only logarithmically on $\gpmx$ for $s_{\rm p}=2$,
the shift of the  $\bh$  and $\pg$ components to higher luminosities is mainly caused by 
the fact that a larger number of protons satisfies the threshold conditions
for photohadronic interactions with the primary electron synchrotron photons. The small bump that
emerges at $\sim 10^{19}$~Hz is the peak of the proton synchrotron component. This is typically hidden
by the other components but may appear for high $\gpmx$
and hard ($s_{\rm p} \lesssim 2$) proton distributions. Finally, $V_{\rm A, 2}$ is an 
indicative example of a theoretical spectrum that is approximately flat in $\nu F_{\nu}$ units
and spans $\sim 6$ orders of magnitude in energy, with the  $\bh$  emission playing a key role.

The SED of $V_{\rm A, 3}$ is obtained with a harder electron distribution that results in 
$\beta\simeq 0.1$ for frequencies above the synchrotron self-absorption one.  
We find that both photohadronic components
of the SED have higher luminosity when compared to those of $V_{\rm A, 0}$, with the 
harder synchrotron photon spectrum in this case being the reason. 
This can be understood by an inspection of eqs.~(\ref{eq25}) and (\ref{eq25b}), since
changes in the proton loss rate are also reflected to the injection rate of secondaries. 
If $\beta_1 = 0.5$ and $\beta_2=0.1$, we find
\eqb
\frac{\fpg^{(2)}}{\fpg^{(1)}} = \frac{(1-\beta_2)}{\beta_2 (\beta_2+2)}
\frac{\beta_1 (\beta_1+2)}{(1-\beta_1)}\left(\frac{2 \gp}{\gamma_{\rm p, \pg}^{\rm (th)}}\right)^{\beta_2-\beta_1} \approx 8,
\eqe
 for $\gp=\gamma_{\rm p, \pg}^{\rm (th)}$. 
This value agrees with the relative luminosity shift of the $\pg$ component shown in Fig.~\ref{fig7}, which is
$\sim 0.7$ in logarithmic units. Similarly, the ratio of the  $\bh$  loss rates for two different spectral indices is
\eqb
\frac{\fbh^{(2)}}{\fbh^{(1)}} = \frac{1-\beta_2}{1-\beta_1} \frac{I(\gp, \beta_2)}{I(\gp, \beta_1)}
\left(\frac{2 \gp}{\gamma_{\rm p,   \bh }^{\rm (th)}}\right)^{\beta_2-\beta_1} \approx 1.4\times 10^{-0.4 c_1 -0.25 c_2},
\eqe
where we used $\gp = \gamma_{\rm p,   \bh }^{\rm (th)}$ and approximated $\log I(\gp, \beta)$ with
a second order polynomial of $\beta$, for a fixed $\gp$; details can be found in the appendix.
In the above, $c_1 <0$, $c_2>0$ are the constants of the polynomial that depend on $\gp$,  not strongly though.
For $\gp\simeq 10^5$, the fitting of eq.~(\ref{app-eq1}) results in $c_1  = -0.6$ and $c_2=0.01$. If we substitute these values into
the equation above we find  $\fbh^{(2)}/\fbh^{(1)} \approx 2.5$, which is in rough agreement with the increase
found numerically (see Fig.~\ref{fig7}).
\subsubsection*{Model~B variants}
The variants of Model~B are summarized below:
\begin{itemize}
\item $V_{\rm B, 0}$ for $\gamma_{\rm e, \min}=3\times 10^2$ and  $s_{\rm e}=2.5$
 \item $V_{\rm B, 1}$ for $\gamma_{\rm e, \min}=3\times 10^2$, $s_{\rm e}=3.0$ 
 \item $V_{\rm B, 2}$ for $\gamma_{\rm e, \min}=10^3$, $s_{\rm e}=2.5$ 
 \item $V_{\rm B, 3}$ for $\gamma_{\rm e, \min}=10^3$, $s_{\rm e}=3.0$ 
\end{itemize}
with $V_{\rm B,0}$ corresponding  to the baseline model.
By changing $\gamma_{\rm e, \min}$ and $s_{\rm e}$ we are able to test the way
different peak frequencies and spectral indices of the synchrotron spectra
affect the contribution of the  $\bh$  and $\pg$ components to the SED.
The respective photon spectra are shown in Fig.~\ref{fig5}.
First, let us compare variants with different $s_{\rm e}$ and fixed $\gamma_{\rm e, \min}$, i.e. 
$V_{\rm B, 0}-V_{\rm B,1}$ and $V_{\rm B,2}-V_{\rm B, 3}$.
On the one hand, we find that softer synchrotron spectra result in a luminosity decrease
of the $\pg$ component, which is directly related to the decrease of
$\fpg$ (see right panel of Fig.~\ref{fig1}).
On the other hand, the luminosity of the  $\bh$  component either decreases (black lines)
or remains approximately constant ( blue lines) as the synchrotron spectra become softer, which
is related to the fact that the dependence of $\fbh$ on $\beta$
differs significantly between protons with different Lorentz factors-- see also right panel of Fig.~\ref{fig1}.
Next we compare variants of the model with different $\gamma_{\rm e, \min}$ and fixed $s_{\rm e}$, i.e. 
$V_{\rm B,0}-V_{\rm B,2}$ and $V_{\rm B,1}-V_{\rm B,3}$. 
An increase of $\gamma_{\rm e, \min}$ by a factor of 3 shifts
the peak frequency by $\sim$ one order of magnitude, and at the same time leads to an increase of both
the  $\bh$  and $\pg$ luminosities. The relative increase of the luminosity is larger for
softer spectra, i.e. for $\beta=s_{\rm e}/2=1.5$. 
Equations (\ref{eq25}) and (\ref{eq25b}) show that both energy loss rates depend on the peak frequency $\nu_{\rm s}$
as $f_{\rm i} \propto \nu_{\rm s}^{-1} \lambda(\beta, \es) \left(\gamma_{\rm p, i}^{\rm (th)} \right)^{-\beta}$, where
$\lambda(\beta, \es)$ is defined in eq.~(\ref{eq24}). 
Since $\lambda$ is in good approximation independent of $\es$ for $\beta >1$ and 
$\gamma_{\rm p, i}^{\rm (th)} \propto \nu_{\rm s}^{-1}$ (see eqs.~(\ref{eq3}) and (\ref{eq4}))
we find that $f_{\rm i} \propto \nu_{\rm s}^{\beta-1}$, which explains the larger
luminosity increase of both components as $\beta$ becomes larger.


\subsubsection{Neutrino emission}
\begin{figure}
\includegraphics[width=0.45\textwidth]{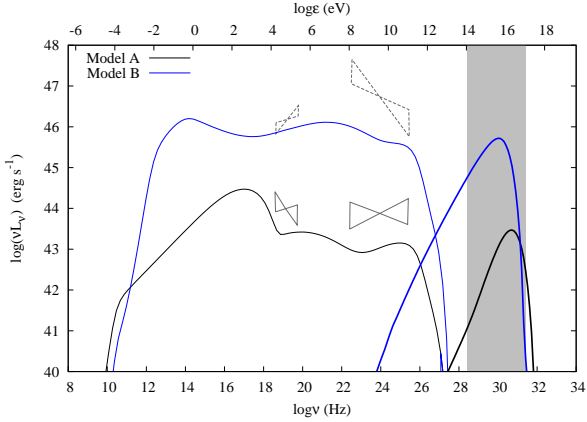} 
\caption{Combined photon (thin lines) and $\nu_{\rm e}+\nu_{\mu}$ (thick lines) spectra obtained for Models A (black lines) and B (blue lines). 
The grey colored region indicates the 0.1-100 PeV energy range. For comparison reasons, 
the average BAT and LAT luminosities for the first (solid lines) and third (dashed lines) BAT luminosity 
bins as defined in  Sambruna et al. (2010) are also shown with bowties. }
\label{fig8}
\end{figure}

As already noted in \S\ref{remarks}, there is a direct link between the $\gamma$-ray and
neutrino emission expected from a BL~Lac object within our scenario. 
Figure \ref{fig8} shows the combined photon and neutrino ($\nu_{\rm e}+\nu_{\mu}$) spectra obtained for the baseline Models A and B.
The grey colored region marks the 0.1-100 PeV energy range and the bowties corresponding to the average BAT and LAT luminosities  \citep{sambruna10}
are overplotted for comparison reasons. 
A few things that are worth commenting follow:
\begin{itemize}
 \item the energy separation $R_{\nu, \gamma}$ of the neutrino and synchrotron from $\pg$ pairs components is
 in agreement with eq.~(\ref{eq8c}), at least for Model~A. 
 When applied to Model~B,  eq.~(\ref{eq8c}) results in approximately three orders of magnitude larger
 separation in energy than the one depicted  in Fig.~\ref{fig8}. 
 The reason is that eq.~(\ref{eq8c}) has been derived under the assumption that there are protons
 energetic enough to satisfy the threshold condition for $\pg$ interactions with photons having energy $\es$.
 In Model~B, however, protons with $\gp = \gpmx$ (see Table~\ref{tab0}) do not satisfy this condition.
 By inverting eq.~(\ref{eq3})  and setting $\gamma_{\rm p, \pg}^{\rm (th)}=\gpmx$, we find that the threshold
 condition for photopion production is satisfied for $\nu \gtrsim 10^{17}$~Hz, which explains the derived
 value of $R_{\nu, \gamma}$.
 \item the neutrino peak energy lies in the $1-100$~PeV energy range in agreement with eq.~(\ref{eq8b}).
 To estimate the peak neutrino energy for Model~B, one has to replace $\nu_{\rm s}$ in eq.~(\ref{eq8b}) with $\sim 10^{17}$~Hz, for the same
 reason explained above. Since $\delta\sim 10$, the peak energy of the neutrino spectrum is expected
 to be $\gtrsim 20$~PeV (see eq.~(\ref{eq8b})), unless the peak of 
 the synchrotron component shifts to $\nu_{\rm s} \gtrsim 10^{17}$~Hz. In this regard, HBLs favour
 the production of neutrinos with a few PeV energy.
 \item in both models, which are described by very different parameters,
 the  $\bh$ , $\pg$ and neutrino components of the blazar emission have comparable luminosity. Our scenario establishes 
 a connection not only between the observed $\gamma$-ray flux 
 and the expected neutrino flux  from  a BL~Lac object, but also links the hard X-ray/soft $\gamma$-ray flux with both of them.
 We caution the reader that application of the model to specific sources, may lead to parameter values that suppress
 the $\bh$  emission, and thus leading to $L_{\rm   \bh } \ll L_{\gamma} \sim L_{\nu}$ (see also discussion in \S\ref{remarks}). 
 This is not,  however, the generic case.
 \end{itemize}

\section{Discussion}
\label{discussion}
In the present paper we  examimed the impact of  $\bh$ 
pair production on the leptohadronic model of blazar emission. 
This process has played so far only a secondary role in modelling of MW spectra
basically because the associated loss rate is typically smaller than the one of $\pg$
interactions that act as a competing loss mechanism for protons.
As a first order approximation, one could thus neglect the  $\bh$  process as a proton 
energy loss mechanism. 

This has been indeed the
standard approach in the literature so far, although it was noted (e.g. BSR90)
that there are parameter regimes where the  $\bh$  loss rate becomes comparable or even surpasses
the $\pg$ loss rate. In our effort to reassess the role of  the $\bh$ process in the context of leptohadronic models of blazar emission, 
we compared the photohadronic loss rates for parameter values 
relevant to blazars (Figs.~\ref{fig1} and \ref{fig2}) using 
analytical expressions (eqs.~(\ref{eq15b}), (\ref{eq18}), (\ref{eq25}) and (\ref{eq25b})) 
that reveal the dependence of each loss rate on the various
parameters. For the  $\bh$  loss rate, in particular, we  derived a useful expression for the case 
of a power-law photon target field (eq.~(\ref{eq25b}) and appendix), which 
faciliates quick comparisons for the losses of the two basic channels of photohadronic 
interactions.

Besides its role as an energy loss mechanism for high-energy protons, 
$\bh$  pair production acts, more importantly,  as an injection process 
of highly relativistic electron-positron pairs. These 
cool mainly through synchrotron radiation and, in principle,
leave their radiative signature on the blazar MW spectrum.
Yet, this aspect of  $\bh$  pair production has not attained a lot of attention.
The  $\bh$  process has a distinct secondary production spectrum which is much broader than the
one produced from photopion interactions (see e.g. Fig.~4 in DMPR12). 
In the simplest case, it is expected that
the radiative signatures of secondaries produced through the aforementioned photohadronic processes
will have non-overlapping spectra.    
Therefore, even in cases where $\pg$ interactions dominate
the losses, $\bh$  secondary radiation could still be detectable, as its emission would not be hidden by the more luminous photopion
component. 

We showed that if the parameters of the source are such as to
make the synchrotron emission of $\pg$ pairs to appear in 
the GeV/TeV $\gamma$-ray regime, then the $\bh$  component emerges
in soft $\gamma$-rays (Figs.~\ref{fig6} and \ref{fig4}). 
This is a robust prediction of the leptohadronic model
that can also serve as a an independent test for the existence of ultrarelativistic protons
in blazar jets. We argued 
that the `smoking gun' for proton acceleration  in blazar jets 
is not only PeV neutrino emission but also the existence of a third photon 
component which lies between the UV/X-rays produced by primary electrons 
and the GeV/TeV $\gamma$ rays produced by $\pg$ secondaries.

The numerical results presented in Section \ref{photons}
were obtained for parameter sets that differed significantly, yet
they indicated the appearance of a broad $\gamma$-ray
hump with a peak in the sub-GeV regime (Figs. \ref{fig7} and \ref{fig5}).
This is an interesting issue, especially in the light of recent observations that reveal the presence of a wide
high-energy component in LBL spectra that is not easily explained as SSC emission. Typical examples are the
MW spectra of AP Librae \citep{fortin10} and BL~Lacartae \citep{bllac_abdo11}, 
where in addition to the SSC emission, an external Compton (EC) component is
required to explain the broad high-energy component, in a pure leptonic scenario. In principle, 
the MW variability  predicted by the leptohadronic and SSC+EC scenarios
will be different, and this can be used as a diagnostic tool for lifting possible model degeneracies.
We plan to address this issue in a future publication. 

As already noted in \cite{mastetal13}, an important feature of the leptohadronic model
under investigation is that it requires neither ultra-high energy protons ($\gpmx\sim 10^{10}$)
nor strong magnetic fields ($B\gtrsim 20$~G), in contrast to 
the more commonly adopted proton-synchrotron blazar model (e.g. \citealt{aharonian00, mueckeprotheroe01}).
The first statement is a straightforward 
result of eq.~(\ref{eq3}), which
also shows that the proton Lorentz factor at the $\pg$ threshold
and the observed peak frequency of the synchrotron spectrum
are inversely proportional. This implies that as we move from HBLs to LBLs
the required minimum proton energy for pion production on the synchrotron photons increases, and
may even exceed the upper limit imposed by the Hillas condition, i.e.  $\gamma_{\rm H} = eB\rb / \mpr c^2$.
We can use the requirement $\gamma_{\rm p, \pg}^{\rm (th)} < \gamma_{\rm H}$ in order to set
a lower limit on the magnetic field strength that is required by the model
\eqb
B > 9.5 \times 10^{-2} \ {\rm G} \ \delta_1 r_{\rm b, 15}^{-1} \nu_{\rm s, 16}^{-1} (1+z)^{-1},
\eqe
where we made use of eq.~(\ref{eq3}). Thus, even for 
LBL sources with $\nu_{\rm s}\simeq 10^{14}$~Hz, our model does
not require very strong magnetic fields, since the lower limit in this case would be $\sim 10$~G.
The uncertainty introduced by the Doppler factor is small, since
it lies typically in the range $10-50$ (e.g. \citealt{celottifabian98, maraschi99}), while
a larger radius would simply relax this constraint. 
Concluding,  strong magnetic fields ($\gtrsim 20$~G) are not necessary for the model to apply.

Since there is no {\sl a priori} reason to exclude weak magnetic fields from our discussion, such as $B\sim 0.1$~G, SSC emission from
primary electrons becomes relevant. 
For fixed $L_{\rm syn}$, $\nu_{\rm syn}$, $\delta$ and $\rb$, weak magnetic fields favour SSC emission. Combining eqs.~(\ref{eq2}) and (\ref{eq6})
we find that Compton scattering of photons
at the peak of the low-energy component by electrons with $\gamma_{\rm e}$ take place in the Thomson regime, i.e.
$\gamma_{\rm e} \xs = 0.08 (\nu_{\rm s, 16} / \nu_{\gamma, 25} )^{1/2} < 3/4$. The typical frequency of the upscattered photons is then written as 
\eqb
\nu_{\rm ssc} \approx \gamma_{\rm e}^2 \nu_{\rm s} \approx 10^{24} \ {\rm Hz} \ \delta_1^2 (1+z)^{-1} \nu_{\gamma, 25}^{-1}.
\eqe
For low enough magnetic fields the observed $\gamma$-ray emission may be therefore the combined result of synchrotron emission from $\pg$ secondary pairs
and SSC emission of primary electrons (see Fig.~\ref{fig6}). In this regard, the present leptohadronic scenario
simplifies into a pure SSC model, by assuming only low enough values
of the proton injection luminosity.

We constrained our analysis to cases where 
the emission from EM cascades is subdominant, as shown in Fig.~\ref{fig4}. EM cascades
are initiated by the absorption of VHE $\gamma$-rays that are produced by neutral pion decay.
The optical depth for their absorption can be estimated as
\eqb
\tgg\left(\epsilon_{\pi^0 \rightarrow \gamma \gamma}^{\rm (th)}\right) \approx 9\times10^{-4}\frac{L_{\rm syn, 45}}{r_{\rm b, 15} \delta_1^3 \nu_{\rm s, 16}(1+z)},
\label{tgg}
\eqe
where the superscript ``th'' is used as a reminder of the parent proton's energy (see eq.~(\ref{eq3})).
For the derivation of the above, we approximated (i) the cross-section for photon-photon absorption 
as $\sigma_{\gamma \gamma} = 0.625 \sth H(x_{\gamma} x-2)/(x_{\gamma} x)$, where $x$ and $x_{\gamma}$
denote the energies of two arbitrary photons in the comoving frame in units of $\mel c^2$, and (ii) 
the low-energy component of the SED by the
monoenergetic photon distribution defined in  eq.~(\ref{eq12}).  
Only a fraction $\tgg$ of the VHE luminosity $L_{\pi^0 \rightarrow \gamma \gamma} \simeq (1/2) \fpg L_{\rm p}$
will appear in lower energies,  with the ratio of the reprocessed to the total neutrino luminosity being 
approximately given by $\tgg L_{\pi^0 \rightarrow \gamma \gamma}/ L_{\nu} \approx (4/3) \tgg$, for $\tgg <1$.
It is noteworthy that the absorption of VHE $\gamma$-rays from $\pi^0$ decay
as well as the decay of charged pions results in the production of pairs with approximately the same
energy (see also \citealt{petromast12}). Thus, 
the absorbed VHE luminosity will reappear in the same $\gamma$-ray energy
regime where the $\pg$ component lies.
 \section{Summary}
 \label{summary}
 We explored some of the consequences that arise from the presence of 
 relativistic protons in blazar jets. We focused on an often overlooked photohadronic process
of astrophysical interest, namely the  $\bh$  pair production, and 
investigated its emission signatures on the SED of blazars. 
Motivated by the recent progress in high-energy neutrino astronomy \citep{aartsen14}, we adopted
a theorical framework for the MW blazar emission, which associates the $\gamma$-ray flux 
with a high-energy (above a few PeV) neutrino signal. In this context, 
the low-energy hump of the SED is explained by synchrotron radiation
of primary relativistic electrons, whereas the $\gamma$-ray emission is the result
of photopion processes.
After the electron and proton distributions that 
are  necessary for explaining the double humped
blazar SED have been determined, then the  $\bh$  component can be automatically defined, i.e. no additional free
parameters are required. 

We showed that for a wide range of parameters
the synchrotron emission from  $\bh$  pairs fills the gap between the
low and high energy components of the SED, i.e. between hard X-rays ($\gtrsim 40$~keV) and soft $\gamma$-rays ($\lesssim 40$~MeV).
Although its peak luminosity is not always comparable to 
the one emitted in hard $\gamma$-rays, its radiative signature on the blazar
spectrum may still be observable, as it is not hidden from other components.
We demonstrated that the ``$\bh$  bump'' of the SED is a robust prediction of the 
leptohadronic model,
 and as such, may provide
indirect evidence for high-energy neutrino emission from BL~Lac objects with a three-hump SED.
Information is, therefore, required, either from current missions targeting in hard X-rays and soft $\gamma$-rays, e.g. {\sl NuSTAR} \citep{harrison13}
 and {\sl INTEGRAL} \citep{lebrun03, ubertini03}, or from 
future satellites designed for  soft $\gamma$-ray
observations with high sensitivity, such as {\sl PANGU} \citep{pangu14}.

\section*{Acknowledgments}
 We would like to thank the referee Prof. F.~W.~Stecker for his 
useful suggestions.
We would also like to thank Dr.~S.~Dimitrakoudis and G. Vasilopoulos for their
comments on the manuscript.
Support for this work was provided by NASA 
through Einstein Postdoctoral 
Fellowship grant number PF3~140113 awarded by the Chandra X-ray 
Center, which is operated by the Smithsonian Astrophysical Observatory
for NASA under contract NAS8-03060.
\bibliographystyle{mn2e} 
\bibliography{BH.bib}

\appendix
\section[]{Approximate Bethe-Heitler loss rate for a power-law photon distribution}
The proton energy loss rate because of Bethe-Heitler pair production
on an arbitrary isotropic photon field has been presented in B70 (see eq.~(\ref{eq10})), while
CZS92 have approximated function $\phi(\kappa)$ at different $\kappa$-regimes -- see e.g. eqs.~(3.13), (3.14) and (3.18), therein.
Despite the good accuracy of the approximate solution, it is not so useful for the 
analytical manipulation of the integral in eq.~(\ref{eq10}), especially when the integration is to be performed for
a power-law photon distribution ($n(\kappa) \propto \kappa^{-s}$).

\begin{figure}
 \centering
 \includegraphics[width=0.45\textwidth]{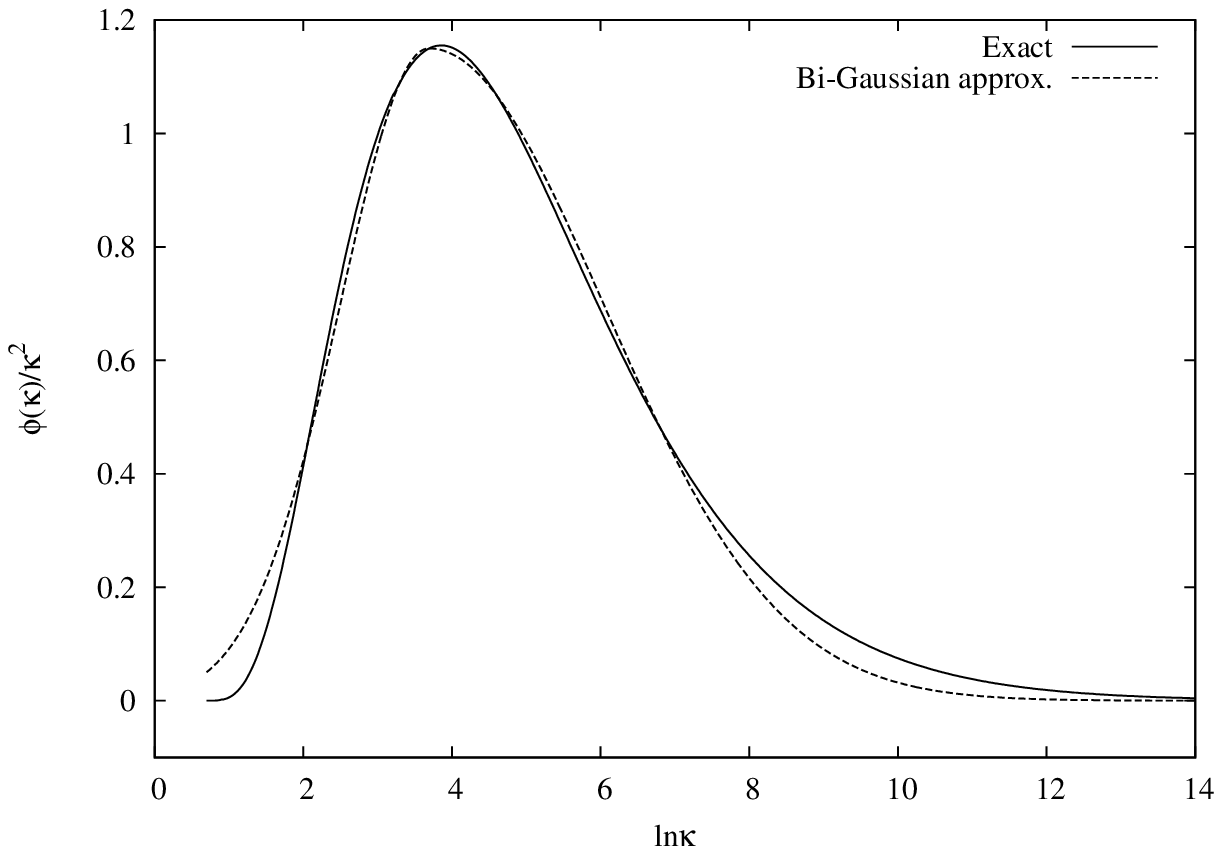}
 \caption{Comparison of the exact (solid line) and bi-Gaussian approximation (dashed line) of the function $f = \phi(\kappa)/ \kappa^2$.}
 \label{app-fig1}
\end{figure}

\begin{figure}
 \centering
 \includegraphics[width=0.45\textwidth]{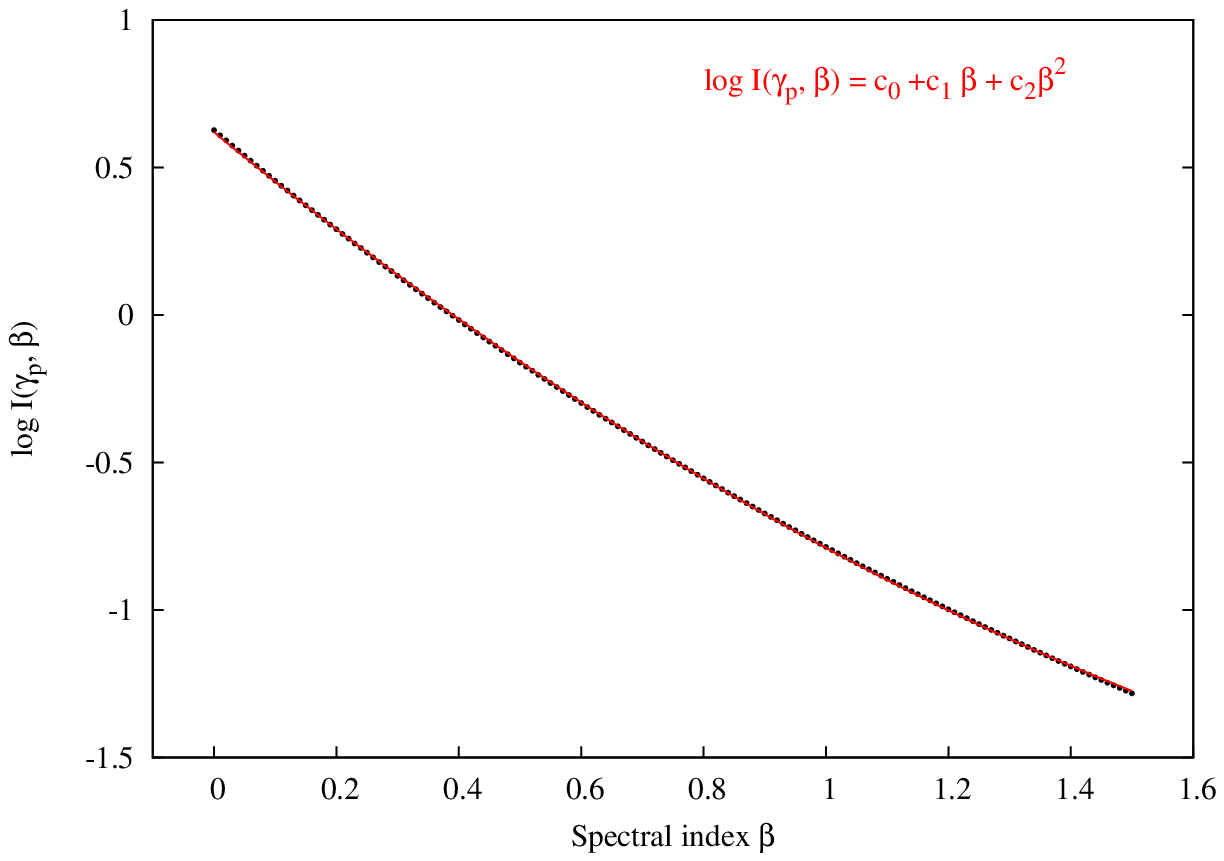}
  \caption{Plot of $\log I(\gp, \beta)$ as a function of the spectral index for $\gp=10^7$ (points).
  The red line is the result of a second order polynomial fit with constants $c_0=0.62$, $c_1=-1.7$ and $c_2=0.29$.
  Other parameters
  used are $x_{\max}=2.7 \times 10^{-5}$ and $x_{\min}=10^{-4} x_{\max}$. }
 \label{app-fig3}
\end{figure}

\begin{figure}
 \centering
 \includegraphics[width=0.45\textwidth]{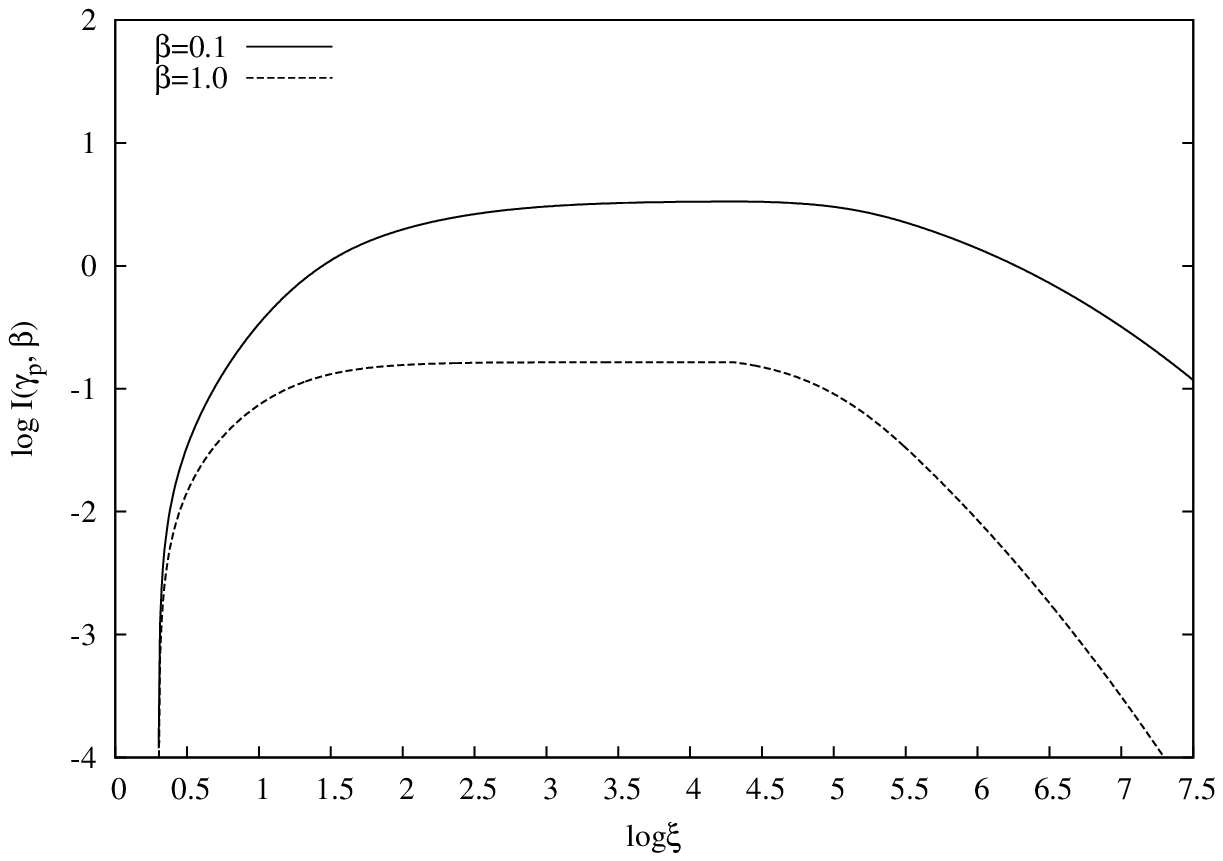}
 \caption{Log-log plot of $I(\gp, \beta)$ as a function of $\xi = 2 \gp/\gamma_{\rm p, BH}^{\rm (th)}$ for $\beta=0.1$ (solid line) and $\beta=1$ (dashed line).
Other parameters   used are $x_{\max}=2.7 \times 10^{-5}$, $x_{\min}=10^{-4} x_{\max}$, and  $\gamma_{\rm p, BH}^{\rm (th)}=4\times 10^4$.}
 \label{app-fig4}
\end{figure}

Here we present a different approximation that facilitates the analytical calculation of $\tbh^{-1}$
for the case of a power-law photon distribution. Instead of focusing on $\phi(\kappa)$, 
we approximate the function $f(\kappa) = \phi(\kappa)/\kappa^2$.
This has a peak of $\simeq 1$ at $\kappa \approx 47$ (see Fig.~2 in CZS92). 
When expressed in terms of $y = \ln \kappa$, function $f$ can be modeled as a bi-Gaussian function
with four free parameters: the position of the peak $y_0$, the standard deviations $\sigma_1$ and $\sigma_2$
of the half Gaussians to the left and to the right of the peak, respectively, and the overall normalization $A$:
\eqb
f(y) = f_0 \left \{ \begin{array}{cc}
                   e^{-a_1 \left(y-y_0 \right)^2}, & y \le y_0 \\
                   e^{-a_2 \left(y-y_0 \right)^2}, & y > y_0,
                  \end{array}
\right.
\label{app0}
\eqe
where $f_0 = A/\sqrt{2 \pi}$ and $a_i = 1/ 2\sigma_i^2$.  For $f_0 = 1.15$, $y_0 = \ln40$, $a_1=0.35$, and $a_2=0.09$,
we find a good agreement with the exact expression in the range $7 \lesssim \kappa \lesssim 5\times 10^3$ (see Fig.~\ref{app-fig1}). For these values, 
the fractional error lies in the range $-5.8\% \le \Delta f/f \le 3.4 \%$ and has a mean value of $\sim 0.1 \%$.
Using the bi-Gaussian approximation for the calculation of the integral in eq.~(\ref{eq10}), for 
the photon distribution of eq.~(\ref{eq21}), we find
\eqb
I_0 = N_0 \int_{y_{\min}}^{y_{\max}} dy e^{-\beta y} f(y)= N_0 I(\gp,\beta),
\label{app-eq0}
\eqe
where $N_0 =\mel c^2 n'_0 (2\gp \xs)^{\beta +1}$, $y_{\min}=\max[\ln2, \ln(2\gp x_{\min})]$ and 
$y_{\max}= \ln (2\gp x_{\max})$. 
The integral can be now performed analytically and results in
\eqb
\frac{I(\gp, \beta)}{f_0 e^{-\beta y_0}} =\left\{ \begin{array}{ll}
                            \!  f_1 erf(x_1)\biggr \rvert_{y_{\min}}^{y_0}\!\! + \!
       f_2 erf(x_2) \biggr \rvert_{y_0}^{y_{\max}}, & y_{\min} \le y_0 \le y_{\max} \\
      \! f_1 erf(x_1)\biggr \rvert_{y_{\min}}^{y_{\max}}, & y_{\max} \le y_0 \\
      \! f_2 erf(x_2)\biggr \rvert_{y_{\min}}^{y_{\max}}, & y_{\min} \ge y_0,
                                         \end{array}
\right.  
\label{app-eq1}
\eqe
where $erf(x)$ is the error function and
\eqb
f_i &  = & \sqrt{\frac{\pi}{4a_i}} e^{\beta^2/ 4 a_i} \\
x_i & = & \frac{\beta + 2a_i (y-y_0)}{2 \sqrt{a_i}} 
\eqe
for $i=1,2$. For a fixed proton energy, the logarithm of $I(\gp, \beta)$ can be modelled
by a second order polynomial of the spectral index, i.e. 
$\log I(\gp, \beta)=c_0+c_1 \beta + c_2 \beta^2$, with the numerical values of the constants $c_i$
depending on $\gp$.
Figure \ref{app-fig3} shows $I(\gp, \beta)$ in logarithmic units, as a function of
$\beta$ (points)  for $\gp=10^7$. The red line is the result
of a non-linear fit with $c_0=0.62$, $c_1=-1.7$ and $c_2=0.29$.
We verified that this result is
not sensitive on the the ratio of the minimum and maximum energies of the
photon distribution. 
The dependence of $I(\gp, \beta)$ on $\xi = 2\gp/\gamma_{\rm p, BH}^{\rm (th)}$, where $\gamma_{\rm p, BH}^{\rm (th)}=4\times 10^4$,
is exemplified in Fig.~\ref{app-fig4} for two values of the spectral index $\beta$.
Interestingly, $I(\gp,\beta)$ is constant for a wide range of $\xi$ values, and starts to 
decrease with $\xi$ only for large enough values (see Fig.~\ref{app-fig4}).

\begin{figure}
 \centering
 \includegraphics[width=0.45\textwidth]{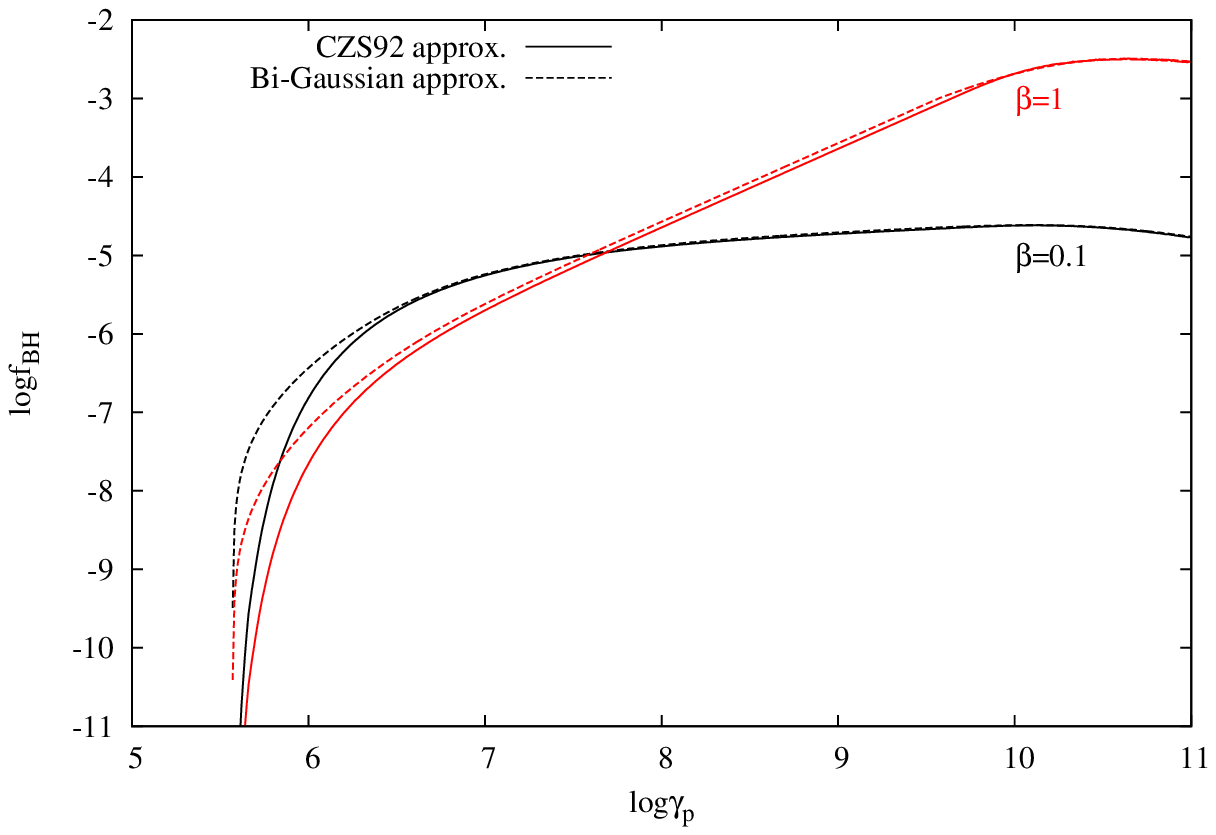}
 \caption{Comparison of $\fbh$ when calculated using the CZS92 (solid line) and bi-Gaussian (dashed line) approximations for a power-law photon distribution with
 $\beta=0.1$ (black lines) and $\beta=1$ (red lines). For the parameters used, see text.
}
 \label{app-fig2}
\end{figure}

The dimensionless loss rates $\fbh$ calculated using the CZS92 and bi-Guassian approximations
for the case of a power-law photon distribution are presented 
in Fig.~\ref{app-fig2}. The parameters used are: $L_{\rm syn}=10^{45}$~erg/s, $\rb=3\times 10^{15}$~cm, $\delta=30$,
$\nu_{\rm s}=10^{16}$~Hz, $\xs=x_{\max}=2.7\times10^{-6}$ and $x_{\min}=10^{-4}x_{\max}$.
We show the results for two spectral indices, i.e. $\beta=1$ (red) and $0.1$ (black). The normalization of the photon distribution 
for $\beta=1$ and $\beta=0.1$ is $n'_0=10^{13.3}(\mel c^2)^{-1}$ and $10^{14.2}(\mel c^2)^{-1}$~cm$^{-3}$ erg$^{-1}$, respectively.
Except for the difference close to the threshold, where the fractional error of the approximation takes its maximum value (see Fig.~\ref{app-fig1}),
the two results are in very good agreement. We verified this for different different values of the ratio $x_{\max}/x_{\min}$ and spectral indices.
\end{document}